\documentclass[twocolumn]{aastex63}
\usepackage[utf8]{inputenc}

\usepackage{enumitem}
\usepackage{amsmath}
\usepackage{physics}
\usepackage{color}
\usepackage{graphicx}
\usepackage{wrapfig}
\usepackage{booktabs}
\usepackage{float}
\usepackage{multirow}
\usepackage{booktabs}
\usepackage{nccmath}
\usepackage{stackrel}
\usepackage{mathtools}
\usepackage{hyperref}

\newcommand{\E}{\mathrm{E}}

\newcommand{\Cov}{\mathrm{Cov}}
\newcommand{\eqset}{\stackrel{\mathclap{\mbox{set}}}{=}}

\begin{document}

\accepted{2022-04-21}
\submitjournal{Astronomical Journal}

\title{Correlated Read Noise Reduction in Infrared Arrays Using Deep Learning}

\author[0000-0002-0997-4827]{Guillaume Payeur}
\affiliation{Department of Physics, McGill University, Montreal, QC, Canada H3A 2T8}

\author[0000-0003-3506-5667]{Étienne Artigau}
\affiliation{Observatoire du Mont-M\'egantic, D\'epartement de Physique, C.P.~6128 Succ. Centre-ville, Montr\'eal, QC H3C~3J7, Canada}

\affiliation{Institute for Research on Exoplanets, Universit\'e de Montr\'eal, D\'epartement de Physique, C.P.~6128 Succ. Centre-ville, Montr\'eal, QC H3C~3J7, Canada}

\author[0000-0003-3544-3939]{Laurence Perreault Levasseur}
\affiliation{Universit\'e de Montr\'eal, D\'epartement de Physique, C.P.~6128 Succ. Centre-ville, Montr\'eal, QC H3C~3J7, Canada}

\affiliation{Mila - Quebec Artificial Intelligence Institute,
Montr\'eal, Canada}
\affiliation{Center for Computational Astrophysics, Flatiron Institute, 162 5th Avenue, New York, NY, 10010, USA}

\author[0000-0001-5485-4675]{René Doyon}
\affiliation{Observatoire du Mont-M\'egantic, D\'epartement de Physique, C.P.~6128 Succ. Centre-ville, Montr\'eal, QC H3C~3J7, Canada}
\affiliation{Institute for Research on Exoplanets, Universit\'e de Montr\'eal, D\'epartement de Physique, C.P.~6128 Succ. Centre-ville, Montr\'eal, QC H3C~3J7, Canada}

\keywords{Astronomy data reduction, Near infrared astronomy, Deep learning}

\begin{abstract}
    We present a new procedure rooted in deep learning to construct science images from data cubes collected by astronomical instruments using HxRG detectors in low-flux regimes. It improves on the drawbacks of the conventional algorithms to construct 2D images from multiple readouts by using the readout scheme of the detectors to reduce the impact of correlated readout noise. We train a convolutional recurrent neural network on simulated astrophysical scenes added to laboratory darks to estimate the flux on each pixel of science images. This method achieves a reduction of the noise on constructed science images when compared to standard flux-measurement schemes (correlated double sampling, up-the-ramp sampling), which results in a reduction of the error on the spectrum extracted from these science images. Over simulated data cubes created in a low signal-to-noise ratio regime where this method could have the largest impact, we find that the error on our constructed science images falls faster than a $1/\sqrt{N}$ decay, and that the spectrum extracted from the images has, averaged over a test set of three images, a standard error reduced by a factor of 1.85 in comparison to the standard up-the-ramp pixel sampling scheme. The code used in this project is publicly available on GitHub \footnote{\url{https://github.com/GuillaumePayeur/HxRG-denoiser}}
\end{abstract}

\section{Introduction}
\label{sec:Introduction}

Large format near-infrared (NIR) detectors are at the heart of major ground- and spaced-based astronomical instruments, the HgCdTe HxRG  being the most common devices. The H2RG-18 is a detector featuring a pixel size of 18 $\micron$ with cut-off wavelengths of either 2.5 or 5 $\micron$ that was specifically developed for the four NIR instruments onboard the James Webb Space Telescope (JWST; \citealt{Gardner2006}): NIRCam, NIRSPec, NIRISS and FGS. These detectors have near perfect quantum efficiency with very low read-out noise (RON) and dark current \citep{Birkmann2018}. The H2RG program stimulated the development of similar larger format detectors: the 4k$\times$4k H4RG-15 \citep{Hall2012} now used widely on ground-based instruments and the H4RG-10 to be used on the Nancy Grace Roman telescope \citep{Mosby2020}.

RON is the main performance limitation of infrared detectors. The H2RG features 4 rows and 4 columns of non-illuminated reference pixels at the periphery of the detector that can be used to monitor the noise during the sequential readout process. While reference pixels can be used effectively to mitigate the effects of correlated noise, state-of-the-art statistical techniques using them are unsuccessful at removing the entirety of the correlated noise (see \citealt{rauscher_improved_2017}) and leave a typical noise floor of $\sim$5 electrons. 

Minimizing RON is of paramount importance for narrow-band imaging and high-resolution spectroscopic observations of faint astronomical objects.  Medium-resolution spectroscopic observations of faint galaxies with JWST NIRSpec is a good example of such forthcoming observations.  Eliminating or minimizing the correlated noise of medium-resolution NIRSpec observations would effectively increase the aperture of JWST.

Recent advances in deep learning have demonstrated the impressive versatility of these advanced methods for a variety of tasks (e.g., \citealt{silver_mastering_2017}, \citealt{esteva_dermatologist-level_2017}, and \citealt{Hezaveh2017}). In the specific field of astrophysics and astronomy, they have also been shown to have broad applicability (see \citealt{venn_lrp2020_2019}, \citealt{hlozek_data_2019}, \citealt{siemiginowska_astro2020_2019} and references therein for a review of recent applications of deep learning in astrophysics and astronomy), in particular for denoising and super-resolution (see, e.g., \citealt{ulyanov_deep_2020}, \citealt{baso_solar_2019}, \citealt{wei_gravitational_2020}, \citealt{YinLiPNAS2021}, and \citealt{Ni2021}). Machine learning, and in particular deep learning, now offers a new way to remove correlated noise and construct science images. 

In this paper, we explore the use of a convolutional recurrent neural network to remove detector noise from datasets using an H4RG array in the low-flux regime. This is primarily driven by the sequential nature of the temporal evolution of the readout noise through an exposure. We use the specific example of the Near Infra-Red Planet Searcher (NIRPS; \citealt{bouchy_near-infrared_2017}), a high-resolution infrared echelle spectrograh based on a H4RG-15 detector, to provide a proof-of-concept of our method that can be applicable to any other instruments operating  a HxRG detector under a low-flux regime. 

The outline of the paper is as follows.
Section \ref{sec:The case for using deep learning} discusses the readout noise plateau problem, and motivates the use of deep learning to solve it.
Section \ref{sec:Methods} presents the architecture of the neural network that was created, and details how it was trained.
Section \ref{sec:Results} presents the results of the tests done to quantify the improvement that was achieved. 
Section \ref{sec:Discussion and Future directions} discusses the results presented in the previous section, and presents possible future directions for machine learning-based methods.

\section{The case for using deep learning}
\label{sec:The case for using deep learning}

\subsection{The Correlated Readout Noise Problem}

\begin{figure}
  \begin{center}
    \includegraphics[width=0.47\textwidth]{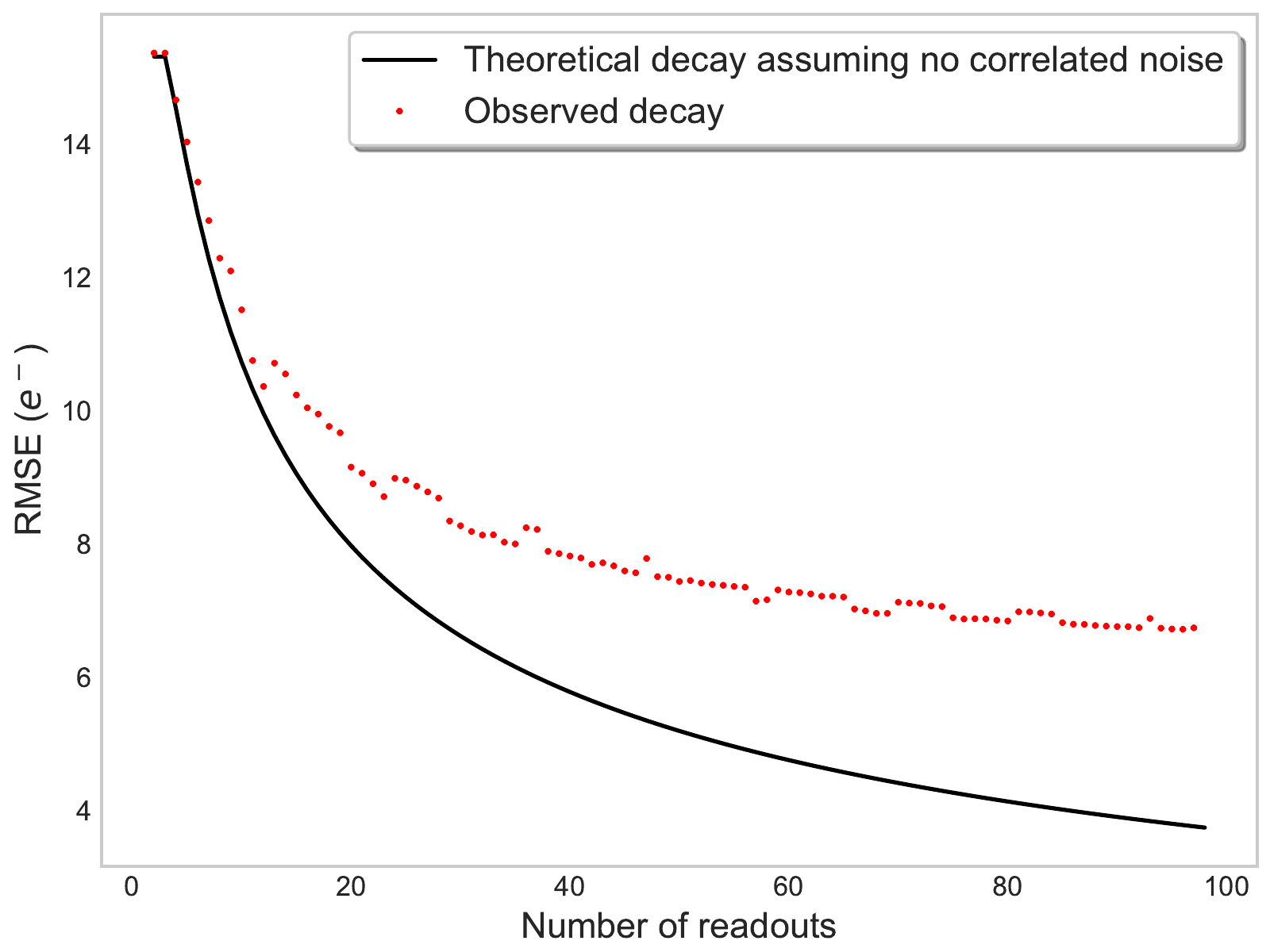}
  \end{center}
  \caption{Observed RON decay on low flux NIRPS data compared to the expected decay in the absence of correlated noise. The observed decay is significantly worse due to the presence of correlated noise in the data.}
  \label{fig:RON fallout}
\end{figure}

With the intent of minimizing readout noise (RON) in science images, near-infrared arrays are read repeatedly, producing a data ramp \citep{fowler_demonstration_1990}. A number of approaches have been used to transform these multiple readouts into a single 2D image representing the scene on the detector. The simplest approach is the correlated-double-sampling (CDS) with a read at the beginning and one at the end of the integration. The science frame is the difference between the last and first frames, effectively removing the common noise, which can, in particular, be due to the reset. A more effective approach on long exposures is to read the detector a number of times at the start of the sequence, thus averaging uncorrelated noise into a first frame, and then repeating the same at the end of the exposure (Fowler sampling). The science frame is then the difference between the mean of frames taken at the end and the mean of frames taken at the start of the exposure. This has the benefit of a smaller RON over the CDS approach, but comes at the cost of a slightly lower effective integration time (photons received during the readouts at the start and end of the sequence have a lower weight in the final difference). Finally, one can perform a linear fit of pixel values as a function of time. In this case, the slope corresponds to the rate of accumulation of photons (i.e. the scene flux), while the intercept corresponds to the effective bias of the detector.  In theory, with N being the number of readouts, the readout noise should fall approximately as $1/\sqrt{N}$ (see appendix \ref{appendix}) for ramp-fitting algorithms. However, the observed trend across all infrared arrays is significantly worse (see Figure \ref{fig:RON fallout}). The RON falls slower than it should in the absence of correlated noise, and nearly plateaus after a certain number of readouts.

\begin{figure*}
  \begin{center}
    \includegraphics[width=1\textwidth]{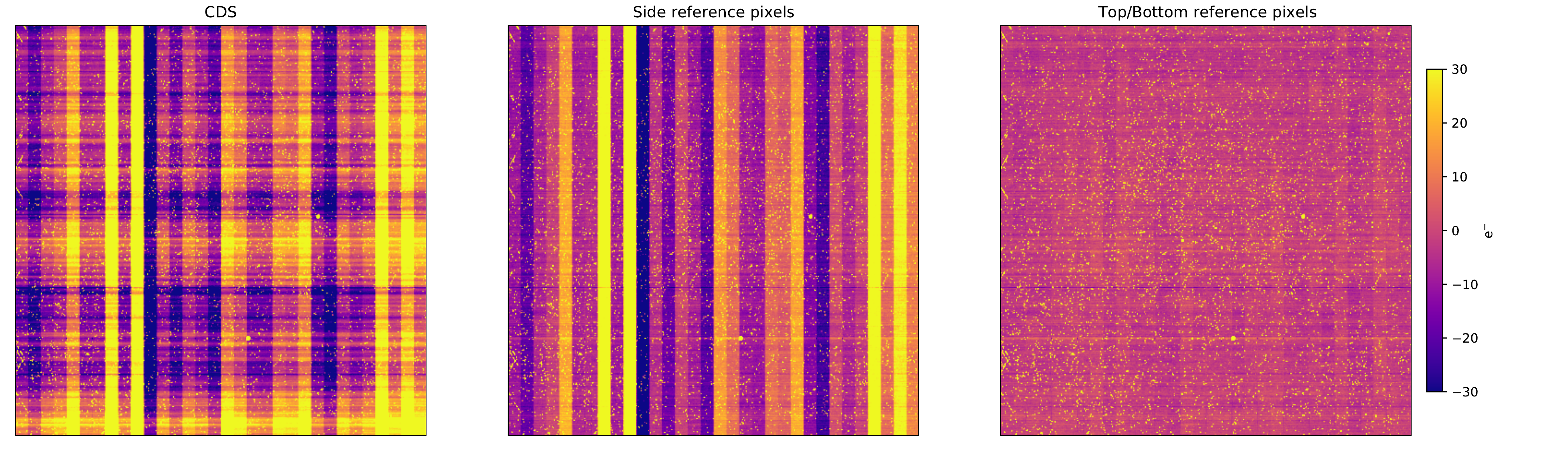}
  \end{center}
  \caption{[Left] An unprocessed dark CDS captured by a NIR array on an H4RG detector. Correlated noise is present in the form of horizontal bands. The standard deviation in pixel value is $\sim28e^-$. [Center] The CDS after it has been corrected using left and right reference pixels. [Right] The CDS after top and bottom reference pixels were used to match amplifier levels. The standard deviation in pixel value is $\sim17e^-$. Correlated noise structures are still visible, showing that current methods are unsuccessful at removing the entirety of the correlated noise.}
  \label{fig:CDS}
\end{figure*}

The cause of the RON plateau can easily be identified. Consider a CDS collected on an H4RG array, as shown in Figure \ref{fig:CDS}. Very clear time-correlated noise structures are present throughout the CDS. The larger part of the noise structures are routinely removed from the data using the array's reference pixels or the array's unilluminated pixels. The result of this process is also shown in Figure \ref{fig:CDS}. Remnants of the time-correlated noise structures can be observed on the corrected CDS, and they participate in creating the readout noise plateau discussed above. It is currently not known how remnants of time-correlated noise structures may be completely removed from the data. The presence of time-correlated readout noise represents a fundamental limitation to the quality of observations in the detector noise limited regime. The purpose of the procedure presented in this paper is to address this problem.

\subsection{Overview of Deep Learning Methods}

In deep learning, neural networks are a class of computational structures designed to identify relationships in input data to produce outputs of interest. The data processing inside such networks is through successive so-called layers, which permit the extraction and analysis of insightful information from data containing complex correlations. Successive layers are composed with each other, meaning that the output of each layer is used as an input to the following one, and the output of the final layer is interpreted as the output of the network. The neural network presented in this paper utilizes two classes of layers: recurrent layers and convolutional layers.

\begin{figure*}
  \begin{center}
    \includegraphics[width=1\textwidth]{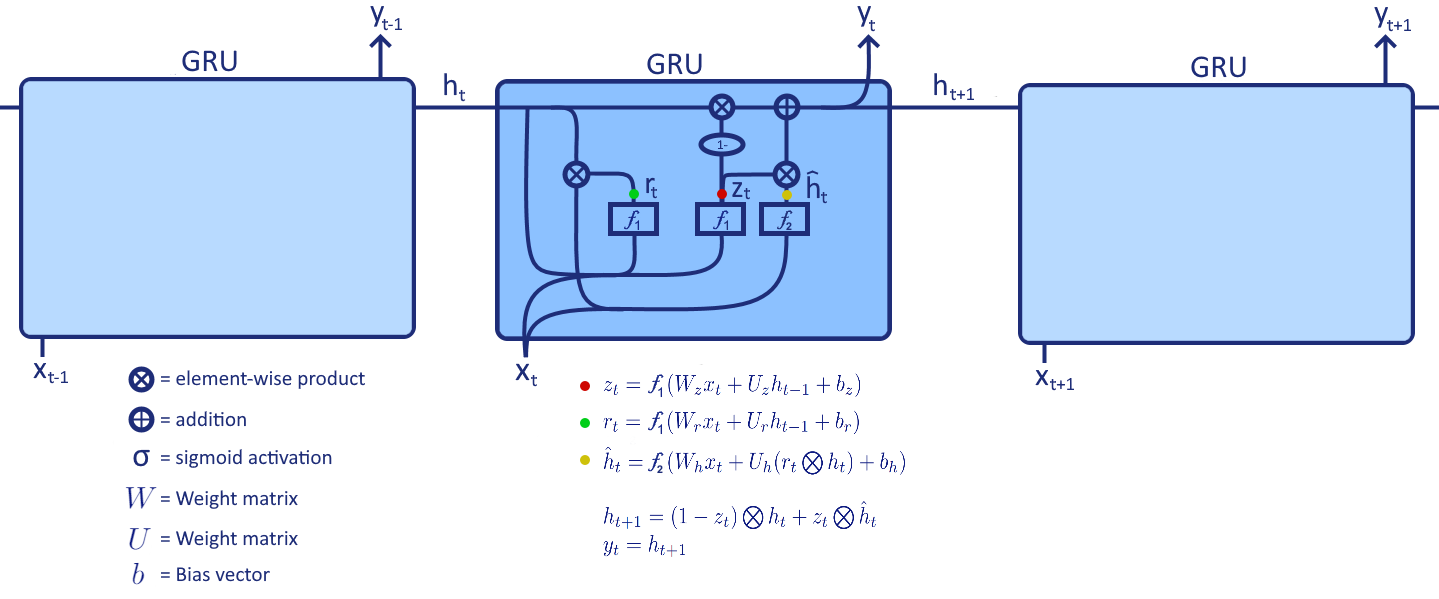}
  \end{center}
  \caption{Diagram of the Gated Recurrent Unit (GRU) cell and layer used in this paper. Each element $x_t$ of a sequence $x$ is processed by the cell along with the hidden state $h_t$ from the previous sequence element, generating an output $y_t$. The outputs $y_t$ form a sequence $y$. The GRU cell may learn through training to extract information from $x$ and embed it into $y$. The so called activation functions $f_1$ and $f_2$ are typically taken to be the sigmoid function and tanh function, respectively.}
  \label{fig:gru}
\end{figure*}

Recurrent layers are specialized in processing sequential data. A recurrent layer consists of a unique cell (see, e.g., Figure \ref{fig:gru}), and each element of a sequence of data is processed individually by this cell in recurrent fashion, producing an output at every step. The so-called hidden state, or memory state, is carried over from processing one element of the sequence to the next, and allows the cell to make use of the information contained in previous elements of the sequence when analyzing the next. This allows the recurrent cell to exhibit dynamic temporal behavior, and therefore makes it particularly well-adapted to analyse time-series data.

Convolutional layers, on the other hand, are predominantly used in computer vision, or image analysis tasks. They convolve their input with a number of filters before applying a non-linear activation function to the resulting maps. 

The values of the convolutional layer filters and the parameters of the recurrent cell in a recurrent layer, which are also commonly collectively referred to as network weights, are optimized during training, a process through which a series of examples for which the true outputs are known is shown to the network. The network weights are then fixed by minimizing a loss function, which quantifies the distance between the truth and the prediction of the network.  Modifying the parameters of the network in the direction opposite to the gradient of the loss improves the performance of the network. This process is called back propagation. For recurrent neural network layers, back propagation is not applicable directly, but its extension, backpropagation through time (BPTT; \citealt{werbos_backpropagation_1990}), achieves the same purpose.
Given a sufficient number of training examples, networks built out of these layers can use these learned parameters to make accurate predictions on previously unseen data.

For instruments using an HxRG detector, while the 2-dimensional science images would see their symmetry structure respected by convolutional layers, it is the temporal evolution of the readout noise throughout an exposure that motivates the use of recurrent layers.
In this work, we present a neural network which is a composition of double-sided recurrent layers and convolutional layers which takes data ramps as input data and produces the flux on pixels of the science image as outputs.

\subsection{Identifying What can be Improved to Motivate Deep Learning}
\label{Motivation Deep Learning}

When a science image is constructed using a NIR spectrometer, the flux on a given pixel is generally not determined solely from data collected on that pixel for multiple reasons, one of which is correlated RON.  Instead, even traditionally, data collected elsewhere on the detector is used as well: reference pixels are one mechanism through which this is conventionally done.  The central idea presented here is to exploit the fact that neural networks are very effective at extracting complex correlations present in data in order to make more extensive use of the noise information contained in other light-sensitive pixels on the images.

Neural networks make it possible to lessen the impact of correlated readout noise in the data significantly. The key factor enabling this is that when computing a prediction for the flux on a given pixel, a neural network may take as input data collected from thousands of neighbouring light-sensitive pixels simultaneously. This vast sample of the RON and of its local evolution through time allows the neural network to better characterize it, and therefore results in a more efficient removal. While the characteristics of the time-correlated RON are not known explicitly, data-driven methods such as neural networks are capable of implicitly inferring such data characteristics from training examples.     

Moreover, computing structures such as convolutional neural networks and recurrent neural networks are especially designed to make use of the spatial and temporal structure of their input data, respectively. For example, a network can easily learn from its training data that neighboring pixels on the detector always receive a closely resembling flux of photons, and ensure that their output images respect this structure. Furthermore, one may expect that noise contributions that are neither spatially nor temporally corrected will average out with the number of readouts. This gain is the motivation behind Fowler and up-the-ramp sampling schemes, but following a removal of correlated noises, one may expect this gain to be even larger.

\section{Methods}
\label{sec:Methods}

This section details the creation and use of the training set, and presents the neural network architecture that was used.

\subsection{Data Ribbons}

\begin{figure*}
  \begin{center}
    \includegraphics[width=1\textwidth]{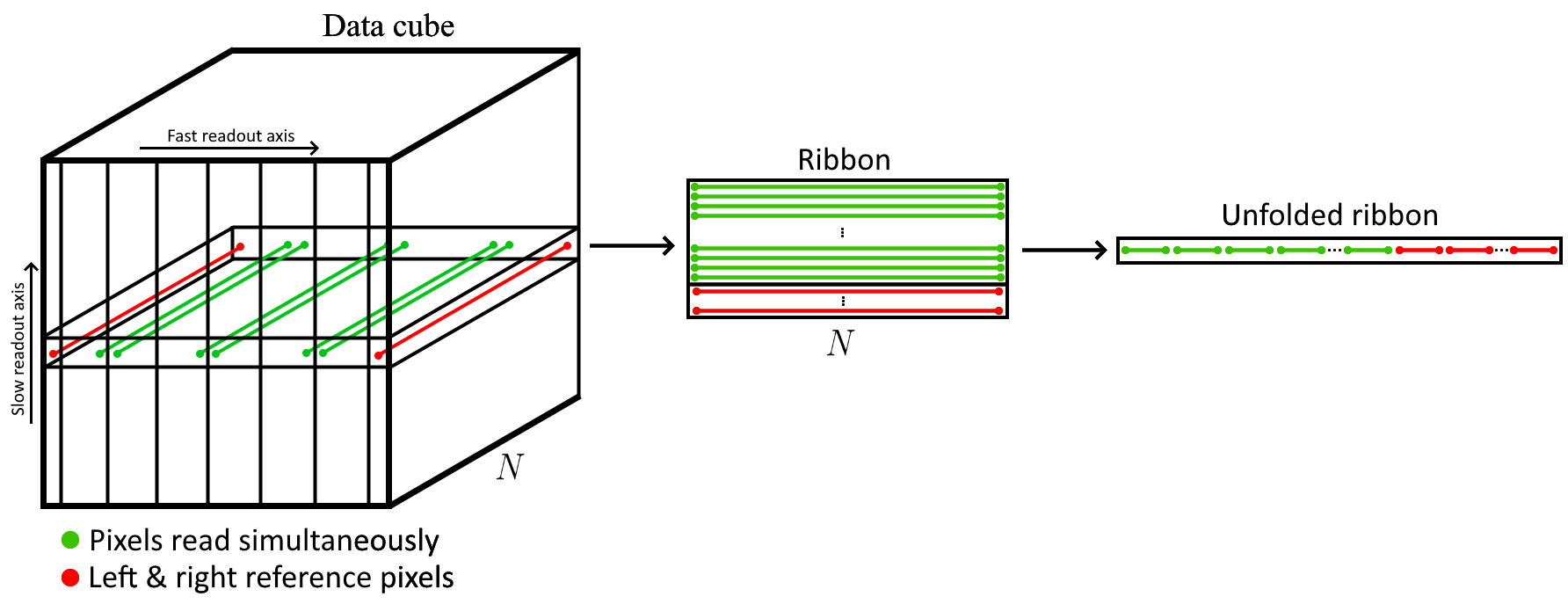}
  \end{center}
  \caption{Representation of our concept of ribbons and unfolded ribbons, and how they are obtained from a data cube of N readouts collected by a NIR array. The ribbon contains all pixels that are read from the detector simultaneously, and the corresponding left and right reference pixels. For visualization purposes, the array is shrunk significantly so that it contains only a few pixels.}
  \label{fig:ribbons}
\end{figure*}

For our purpose, it is desirable to group together pixels of the array which are read simultaneously. To that end, we work with small portions of data cubes which we refer to as ribbons. A ribbon is a collection of all pixels on the array that are read from the detector simultaneously (see Figure \ref{fig:ribbons}). Ribbons also include the corresponding left and right reference pixels, and are reshaped into a two-dimensional array. Ribbons are central to the methods presented here, because it is sequences of unfolded ribbons which are processed as inputs by the neural network, and the unfolded ribbons are treated as elements of a time series.

\subsection{Description and Creation of the Training Dataset}

Training a neural network requires labelled training examples, that is, data for which the underlying astrophysical scene is known. To this end, we created synthetic training data that may be described as partially real, partially simulated, while ensuring that the data is as representative of real data as possible.

To create a synthetic training data ramp, we used optical-design prescriptions of the NIRPS spectrograph, combining echelle order geometry and known fiber size, to project a realistic stellar spectrum, drawn from the  Geottingen spectral library \citep{husser_new_2013}, onto the science array. The order geometry closely reproduced the early tests in the labs. For our simulated frames, we use spectra with $\log$ g=5.0 and solar metallicity. We simulate a Poisson accumulation for that scene, over 100 readouts, and with a peak flux of $\sim$30 ADUs (or equivalently $\sim30 e^-$, as we use a gain very close to one). Concretely, this consists in drawing for 100 frames and every pixel of the array a sample from a Poisson distribution with a rate parameter equal to the per-frame mean flux, peaking at $\sim$0.3 ADU. The 100 frames are added cumulatively to produce 100 readouts. The result is a data ramp containing photon noise. We then add to the synthetic ramp a real NIRPS dark ramp taken in a laboratory, which contains all the expected instrumental noises, including correlated noise. The darks have a read-to-read running difference RMS of $\sim28e^-$, prior to corrections using reference pixels (see, e.g., Figure \ref{fig:CDS}). Before being added to the synthetic data ramp, the darks are corrected by subtracting the mean reference pixel value along rows and columns of the array. This reduces the read-to-read running difference RMS of the darks to $\sim17e^-$. The result of this synthetic procedure is a data ramp that closely resembles real data. Before it is given to the neural network, each readout of the data ramp is further pre-processed in order to eliminate as much correlated noise as possible. For each readout on the data ramp, we subtract the row-wise mean unlit pixel value for each row and the column-wise mean unlit pixel value for each column. We also separate each readout into boxes of 50$\times$50 pixels, and subtract from each box the mean value from all unlit pixels contained in the box. These standard procedures eliminate some of the correlated readout noise at lower frequencies. Additionally, the sequences of readouts collected by hot pixels and dead pixels on the array are replaced by sequences collected by a neighbouring pixel. This is done in order to avoid biasing the neural network's output of neighboring pixels, given its sensitivity to large dynamical ranges.

Using this method, we generate six data ramps of six stars with different darks, and temperatures varying between 3500 K and 5500 K, covering the range of possibilities spanned by our test data. Such variability improves the robustness and the range of applicability of the network.

\begin{figure*}
  \begin{center}
    \includegraphics[width=1\textwidth]{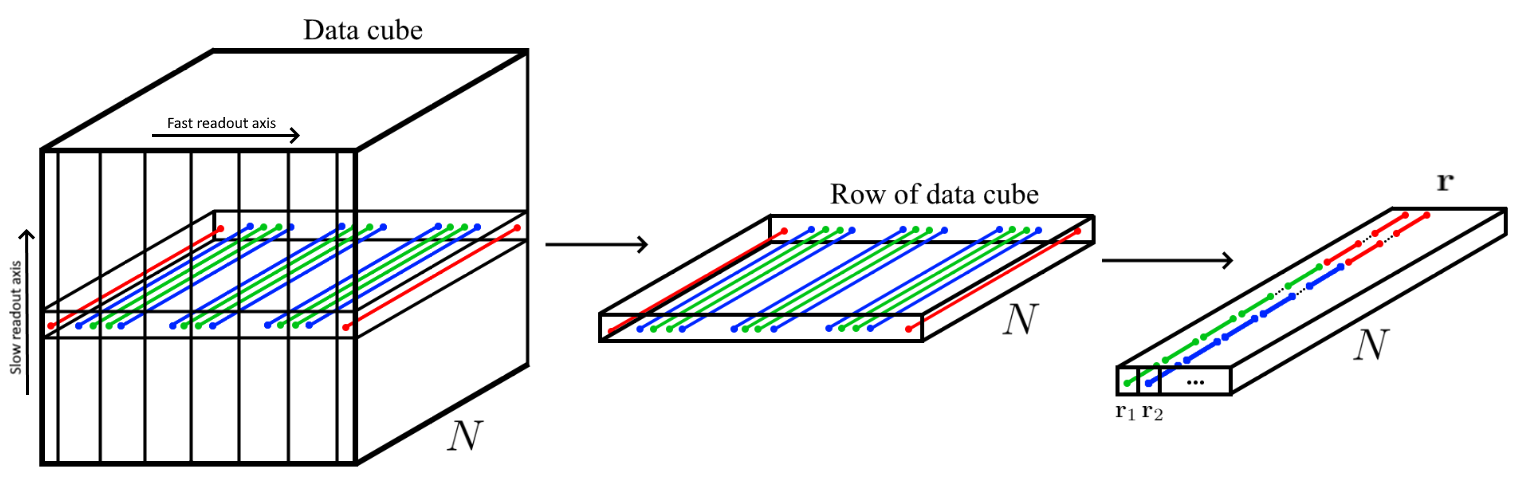}
  \end{center}
  \caption{Creation of a sequence of unfolded ribbons referred to as $\mathbf{r}$ starting from a data cube of N readouts. $\mathbf{r}$ is the concatenation of the unfolded ribbons $\mathbf{r_i}$ contained in the chosen row. The ribbons are ordered as they are read from the detector in order to keep the correlations in the noise intact.}
  \label{fig:inputs}
\end{figure*}

\begin{figure}
  \begin{center}
    \includegraphics[width=0.47\textwidth]{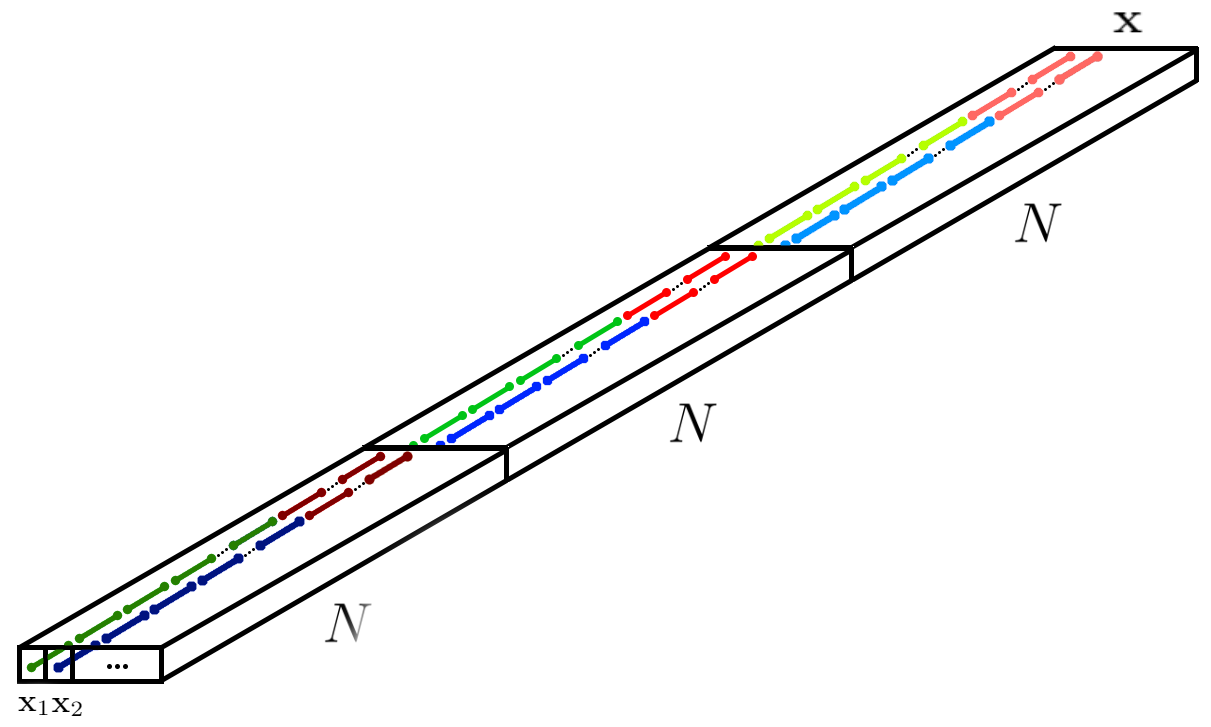}
  \end{center}
  \caption{Creation of the neural network's inputs $\mathbf{x}$ from three consecutive $r$. The inputs $\mathbf{x}$ contain data for three rows of the data cube. The neural network takes in $\mathbf{x}$ as input, and returns predictions for the flux of pixels in the middle row.}
  \label{fig: three rows}
\end{figure}

To allow maximal use of spatial correlations in the images and to allow the network to learn the structure of the noise correlated on all timescales relevant for the reconstruction of the image, ideally entire data cubes would be given as inputs to the neural network. However, the large volume of the cubes would render the model too large for a single GPU and make the training too computationally expensive. As a first proof-of-concept work for this method, we restrict the input data to a single row of the data cube (see Figure~\ref{fig:inputs}). While this still somewhat limits the potential use of spatial structures in the images, the pixels given together as input are the pixels read from the detector along the fast readout axis, and therefore this allows for the neural network to capture time-correlation in the noise along this fast axis. Before it is given as input to the neural network, a row is reshaped such that it is the concatenation of the unfolded ribbons $\mathbf{r_i}$ contained in the row, ordered as they are read from the array in order to keep the correlations in the noise intact (see Figure \ref{fig:inputs}). We call this structure $\mathbf{r}$. The neural network is trained using a set of such $\mathbf{r}$ as training set. However, we find that the performance of the neural network improves significantly if we also provide as input the data taken on the rows above and below the row for which the neural network is generating predictions (see Figure \ref{fig: three rows}). This gives the neural network further context and opportunity to use the spatial correlations in the images, and allows it to capture longer timescale correlations in the readout noise. The neural network is made to only return predictions for pixels in the middle row.

Thus, from the six NIRPS training data ramps, each containing 4088 rows, we get a total of 24528 training examples. We allocate 24028 of them to the training set, while 500 of them taken from random locations on all six training images form the validation set used to fix hyper-parameters and in particular determine the best-performing neural network architecture (see section \ref{subsec:training}).

\subsection{The Neural Network}
\begin{figure}
  \begin{center}
    \includegraphics[width=0.47\textwidth]{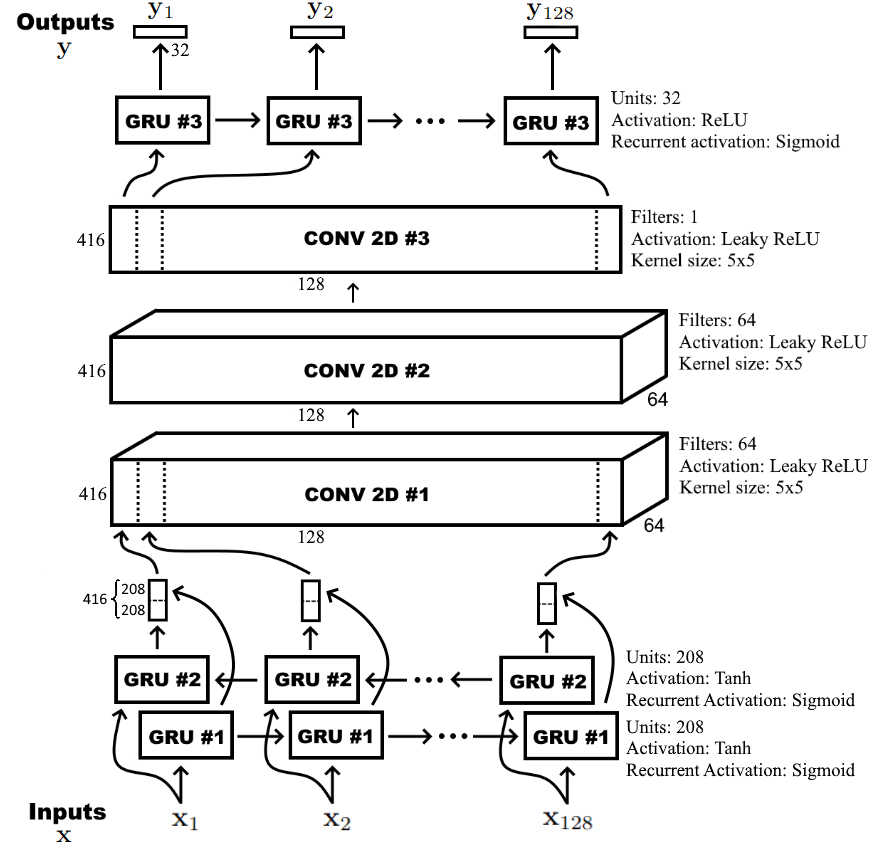}
  \end{center}
  \caption{Architecture of the neural network, and the most important hyperparameters. The numbers surrounding boxes and vectors represent their dimensions. Given as input an array $\mathbf{x}$ containing data collected on three rows of a data cube, it uses a bidirectional GRU layer, followed by three convolutional layers and another GRU layer to generate a point estimate for the flux of pixels contained in the middle row of $\mathbf{x}$, which it returns as $\mathbf{y}$. By generating predictions for each row of a science iteratively, the entire image may be constructed.}
  \label{fig:NN}
\end{figure}

\begin{table*}[]
    \centering
    \caption{GRU and Convolutional Layers Hyperparameters
    }
    \begin{tabular}{ccccc}
        \hline \hline
        \textbf{GRU} & & &\\
        \hline
        Layer & Units & Activation & Recurrent activation & Time orientation\\
        \hline
        1 & 208 & tanh & sigmoid & forward\\
        2 & 208 & tanh & sigmoid & backward\\
        3 & 32 & ReLU & sigmoid & forward\\
        \hline
        \textbf{CONV 2D} & & &\\
        \hline
        Layer & Number of filters & Activation & Kernel size\\
        \hline
        1 & 64 & leaky ReLU (leak=0.1) & 5x5\\
        2 & 64 & leaky ReLU (leak=0.1) & 5x5\\
        3 & 1 & leaky ReLU (leak=0.1) & 5x5\\
        \hline
    \end{tabular}
    \tablecomments{The number of units of a GRU cell as presented in figure \ref{fig:gru} is the dimension of its hidden state. The activation function and recurrent activation functions are the functions $f_1$ and $f_2$ as shown in figure \ref{fig:gru}, respectively.}
    \label{tab:hyperparams}
\end{table*}

The neural network returns a point estimate for the flux on the pixels of the science image. That is, it returns a single number for each pixel. It is composed of two recurrent layers which are GRU, and three 2D convolutional layers. See Figure \ref{fig:NN} for a diagram of the neural network's architecture and table \ref{tab:hyperparams} for a tabulation the hyperparameters of these layers. 

When a chosen $\mathbf{x}$ is given as input to the first GRU layer, it begins by processing $\mathbf{x_1}$ using its cell. It constructs a hidden state based on that, and moves on to $\mathbf{x_2}$, where the cell is now able to make use of the updated hidden state. It updates the hidden state before it moves on to the next ribbon. The process continues recursively for all $\mathbf{x_i}$. H4RG-15 detectors have 128 pixels per amplifier on the fast readout axis, so that in our application to NIRPS the sequence ends at $\mathbf{x_{128}}$. At each step, the GRU cell produces an output vector. Consequently, the GRU layer as a whole produces an ouput sequence. The first recurrent layer is bidirectional, meaning that it has two cells, processing the data in forward and reverse temporal directions. At the end, the outputs for each time step are concatenated, and given as input to the first convolutional layer.

The first convolutional layer receives as input a two-dimensional array which it treats as an image. It uses a set of 5$\times$5 filters, and convolves each one with the image. It concatenates the output of each convolution and passes the resulting three dimensional array to the next convolutional layer. The process is repeated with the second convolutional layer. Having only 1 filter, the third convolutional layer reduces the three-dimensional array it receives to a two-dimensional array again, before it passes it to the last GRU layer. For all convolutional layers we use zero padding to conserve the size of the arrays.  

The last GRU layer is very similar to the first, except that it is not bidirectional. It returns as a sequence $y$ the neural network's predictions for the flux values of all pixels in the middle data cube row of $\mathbf{x}$. H4RG-15 detectors have 32 amplifiers of 128 pixels, so that in our application to NIRPS the output is a sequence of 128 vectors of dimension 32. Each element $y_i$ in that sequence contains the predictions for the pixels contained in $x_i$. As a whole the sequence $y$ contains the predicted flux values for one row of the array. To construct a full science image, we proceed row by row, and combine the rows afterwards.

\subsection{Training}
\label{subsec:training}

The training of the neural network is done using the deep learning API Keras \citep{chollet2015keras} running on top of the machine learing platform TensorFlow \citep{tensorflow2015-whitepaper}

In order to train the neural network, we provide it the labels of the training set, which are the flux values of the underlying scene from which an image has been created, and perform back propagation, using the loss function
\begin{align}
    L4 = \frac{1}{N}\sum_{i=1}^{N}\Big(y_{\text{i}}^{\text{scene}} -y_{\text{i}}^{\text{pred}}\Big)^4,
\end{align}
where $i=(1,\dots,N)$ are all the pixels contained in a training example, $y_{\text{i}}^{\text{scene}}$ is the true value for the flux on that pixel as given by the scene, and $y_{\text{i}}^{\text{pred}}$ is the prediction of the neural network for the flux on this same pixel. We use the L4 loss rather than the more conventional L2 loss because we find that it helps the neural network learn to reconstruct sharp spectral features more accurately. We attribute this to the fact that spectral features are rare in the training set, hence acting as outliers. The L4 loss is more sensitive to outliers, giving more weight to spectral features during training. After experimenting with losses L2, absolute L3, L4, absolute L5 and L6, we found that L4 leads to the best reconstruction of spectral featuers. We use the Adam optimizer \citep{kingma_adam_2017} with a learning rate of $2\text{x}10^{-3}$ and a decay rate of 0.98 applied every epoch. We train for 50 epochs, each using all 24028 training examples and a batch size of 16. In order to decrease overfitting, during training we use dropout on the weights of the recurrent layers' cells, with a dropout rate of 5\%.

\section{Results}
\label{sec:Results}

\subsection{Results on the Test Set}

Once the training of the neural network is completed, we test its performance on new simulated data, which has not been used in training. We generated three additional test images through the same pipeline as the training set (see section \ref{sec:Methods}), but using different simulated spectra, dark ramps and temperatures. For images 2 and 3, we pick temperatures at either edge of the temperature range used during training, as a mean to demonstrate its versatility. We again have access to the scene underlying the images, and we use it to assess the accuracy of the neural network predictions. We also compare the results with the traditional ramp fit method.

\begin{figure*}[!htbp]
  \begin{center}
    \includegraphics[width=0.99\textwidth]{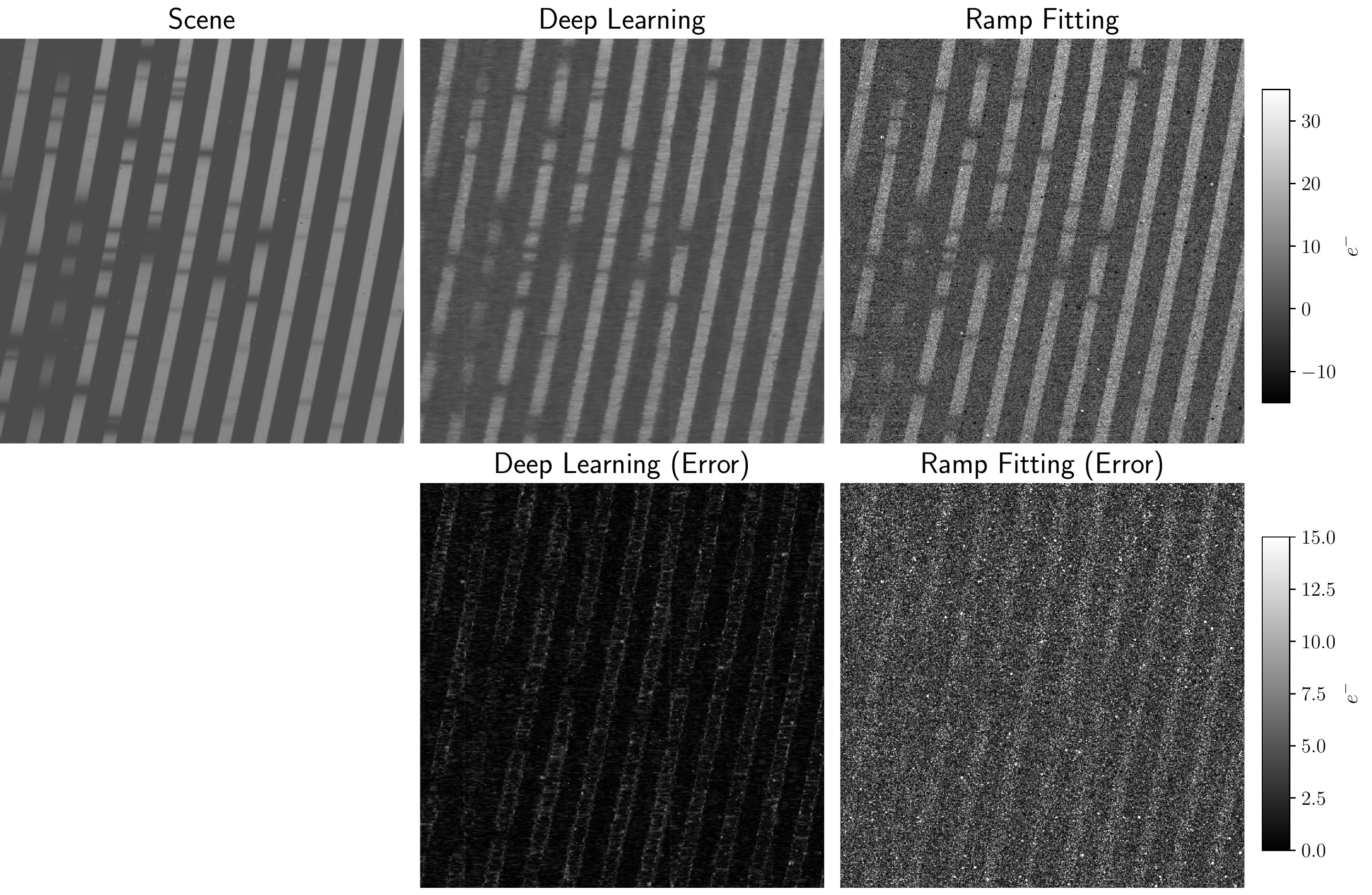}

  \end{center}
  \caption{[Top row] 2D comparison between one of the test science images and its reconstruction done using the neural network and ramp fitting. A region of 500$\times$500 pixels is displayed. The image constructed by the neural network is significantly closer to the scene than is the image constructed by ramp fitting. It is also smoother than the image constructed by ramp fitting, due to the use of spatial correlations by the neural network. [Bottom row] 2D comparison between the absolute error on the reconstruction of the image done using the neural network and ramp fitting. The same region of 500$\times$500 pixels is displayed}
  \label{fig:2D Comparison}
\end{figure*}

\begin{figure}[!htbp]
  \begin{center}
    \includegraphics[width=0.47\textwidth]{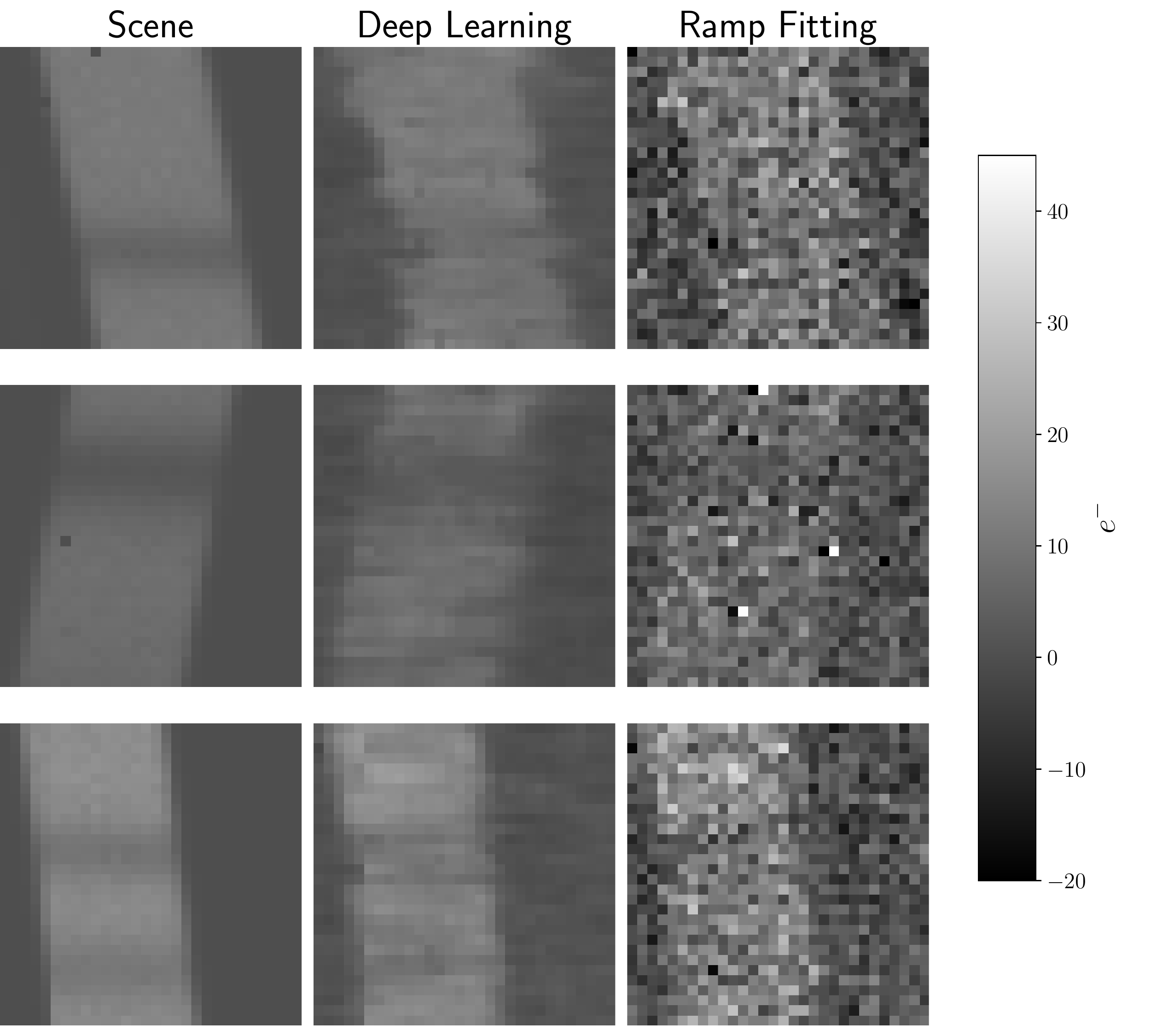}
    
  \end{center}
  \caption{2D comparison between one of the test science images and the reconstruction of one of the test science images done using the neural network and ramp fitting. Three regions of 30$\times$30 pixels are displayed. The image constructed by the neural network is significantly closer to the scene than is the image constructed by ramp fitting.}
  \label{fig:2D Comparison 2}
\end{figure}

On the test images, the error on the predictions of the neural network is significantly lower than the error on predictions obtained by ramp fitting (see table \ref{tab: Errors}). An overview comparison between one scene in the test set the and predictions obtained from the neural network and ramp fitting is shown in Figure \ref{fig:2D Comparison}. A closer comparison on small portions of the image is shown in Figure \ref{fig:2D Comparison 2}. The image constructed by the neural network is much closer to the truth, and absorption lines are reconstructed correctly. The error on the predictions obtained from the neural network and ramp fitting are compared in Figure \ref{fig:2D Comparison}.

\begin{table*}[]
    \centering
    \caption{Pixel-Level and Spectrum Level RMSE on Test Images
    }
    \begin{tabular}{cccc}
        \hline \hline
        & Image 1 & Image 2 & Image 3\\
        \hline
        Temperature & 4500K & 3500K & 6000K\\
        Pixel-level RMSE -- Neural network ($e^-$) & 1.58 & 1.60 & 1.59\\
        Pixel-level RMSE -- Ramp fitting ($e^-$) & 6.22 & 6.03 & 6.05\\
        Pixel-level RMSE reduction & 3.93 & 3.77 & 3.81\\
        Spectrum-level RMSE reduction & 1.87 & 1.84 & 1.84\\
        \hline
    \end{tabular}
    \tablecomments{We compare the results obtained using our neural network to the results obtained using ramp fitting. We discuss the spectrum-level error reduction later in the text.}
    \label{tab: Errors}
\end{table*}

\begin{figure}
  \begin{center}
    \includegraphics[width=0.47\textwidth]{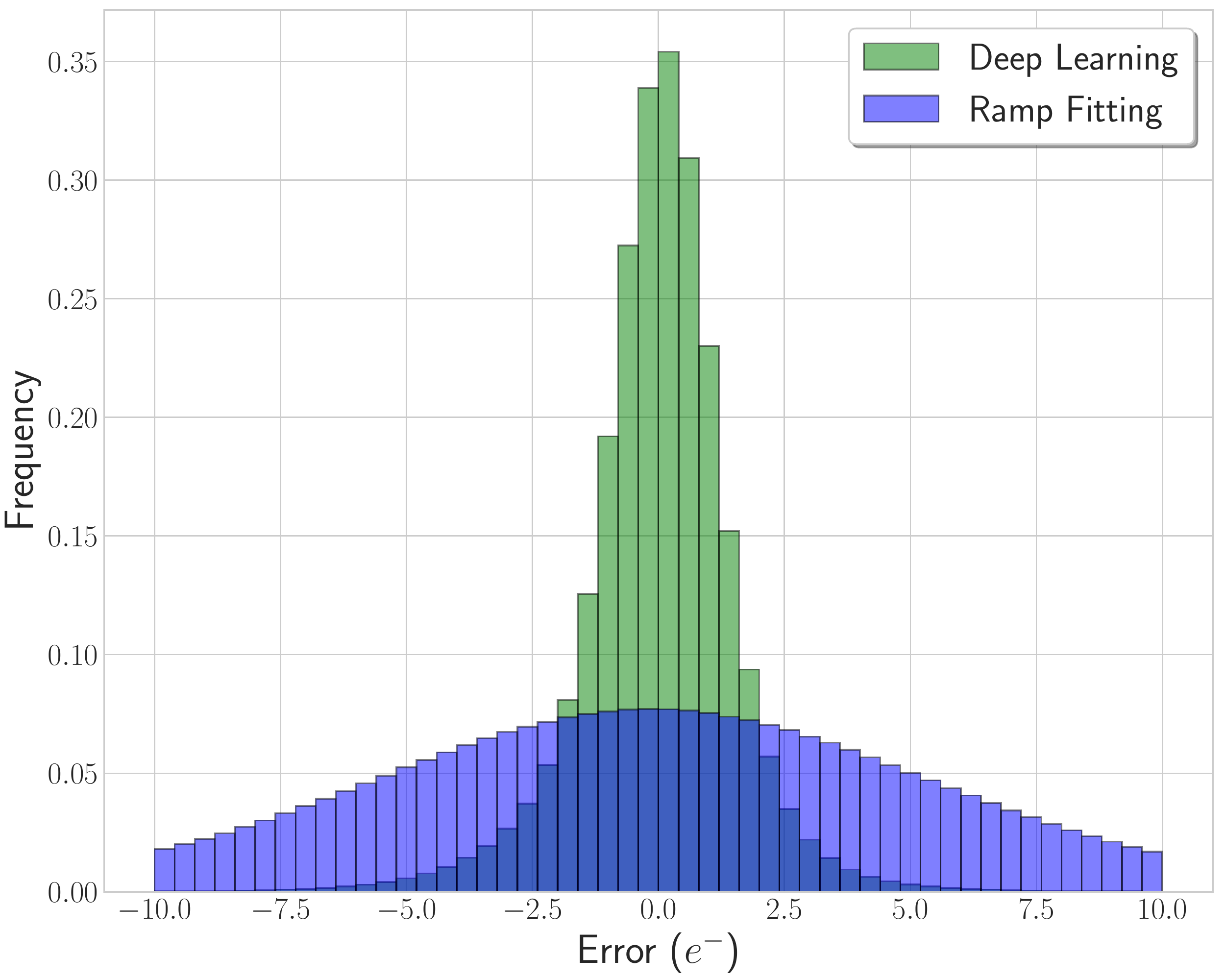}
  \end{center}
  \caption{Distribution of errors on the predictions of the neural network over an entire test image. The distribution of errors resulting from ramp fitting is also displayed. The histogram for the neural network predictions is significantly narrower than that of ramp fitting.}
  \label{fig:Histogram Errors}
\end{figure}

\begin{figure}
  \begin{center}
    \includegraphics[width=0.47\textwidth]{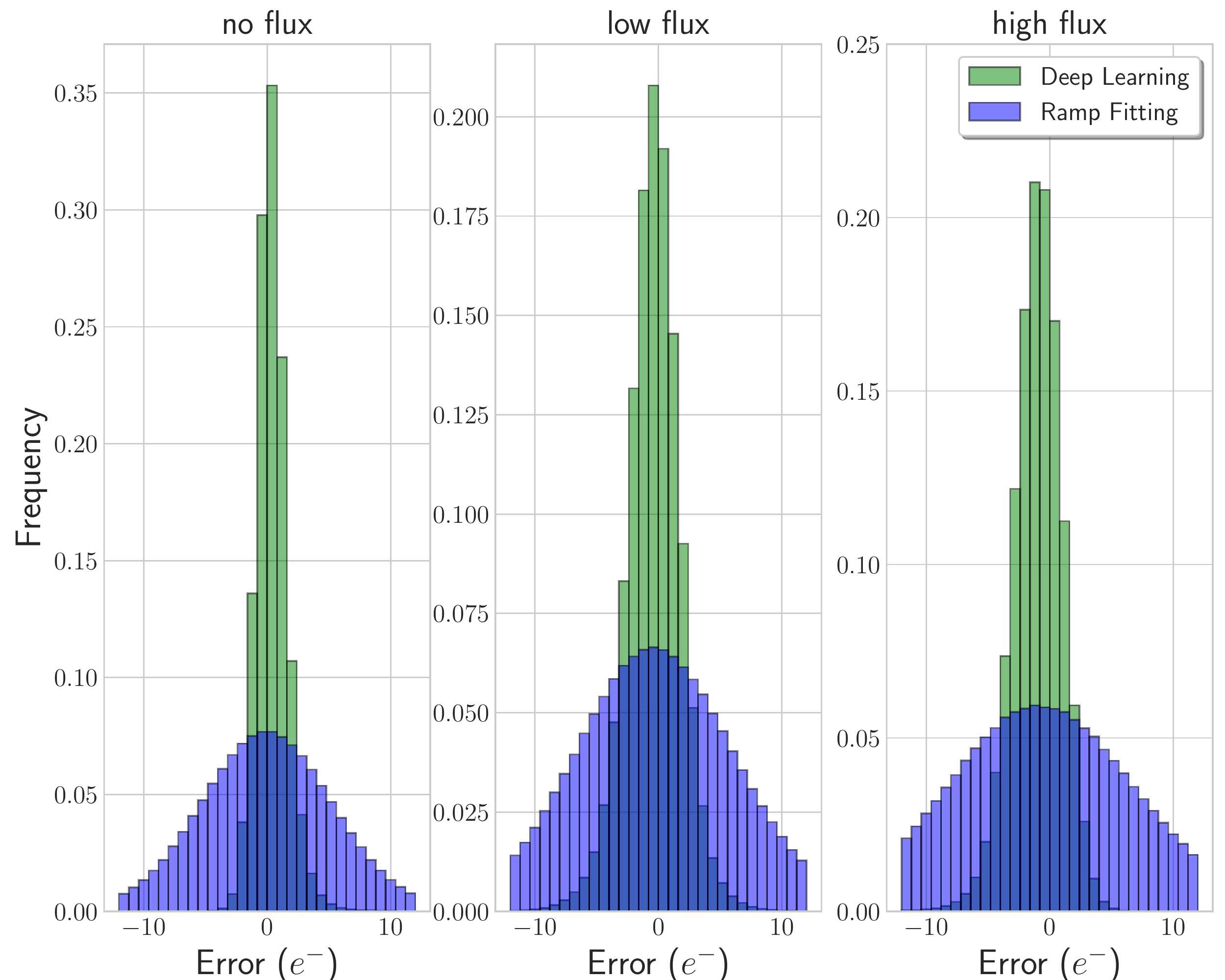}
  \end{center}
  \caption{Distribution of errors on the predictions of the neural network over an entire test image, split into three categories of pixels: no flux ($<1$\,$e^-$), low flux (1\,$e^-$ to 20\,$e^-$), and high flux (20\,$e^-$ to 60\,$e^-$).  
 The distribution of errors resulting from ramp fitting is also displayed. In all cases, the histogram for the neural network predictions is significantly narrower than that of ramp fitting showing that the image constructed using the neural network is closer to the scene than is the image constructed by ramp fitting.}
  \label{fig:Histogram Errors Split}
\end{figure}

The distribution of errors over one of the test images is shown in Figure \ref{fig:Histogram Errors}, and compared to the one obtained by ramp fitting. As an alternative way to view the distribution of errors, Figure \ref{fig:Histogram Errors Split} contains similar histograms, but the pixels on the array are split into three categories: pixels having received no flux (0$e^-$ to 1$e^-$), low flux (1$e^-$ to 20$e^-$), and high flux (20$e^-$ to 60$e^-$).  

\begin{figure*}
  \begin{center}
    \includegraphics[width=1\textwidth]{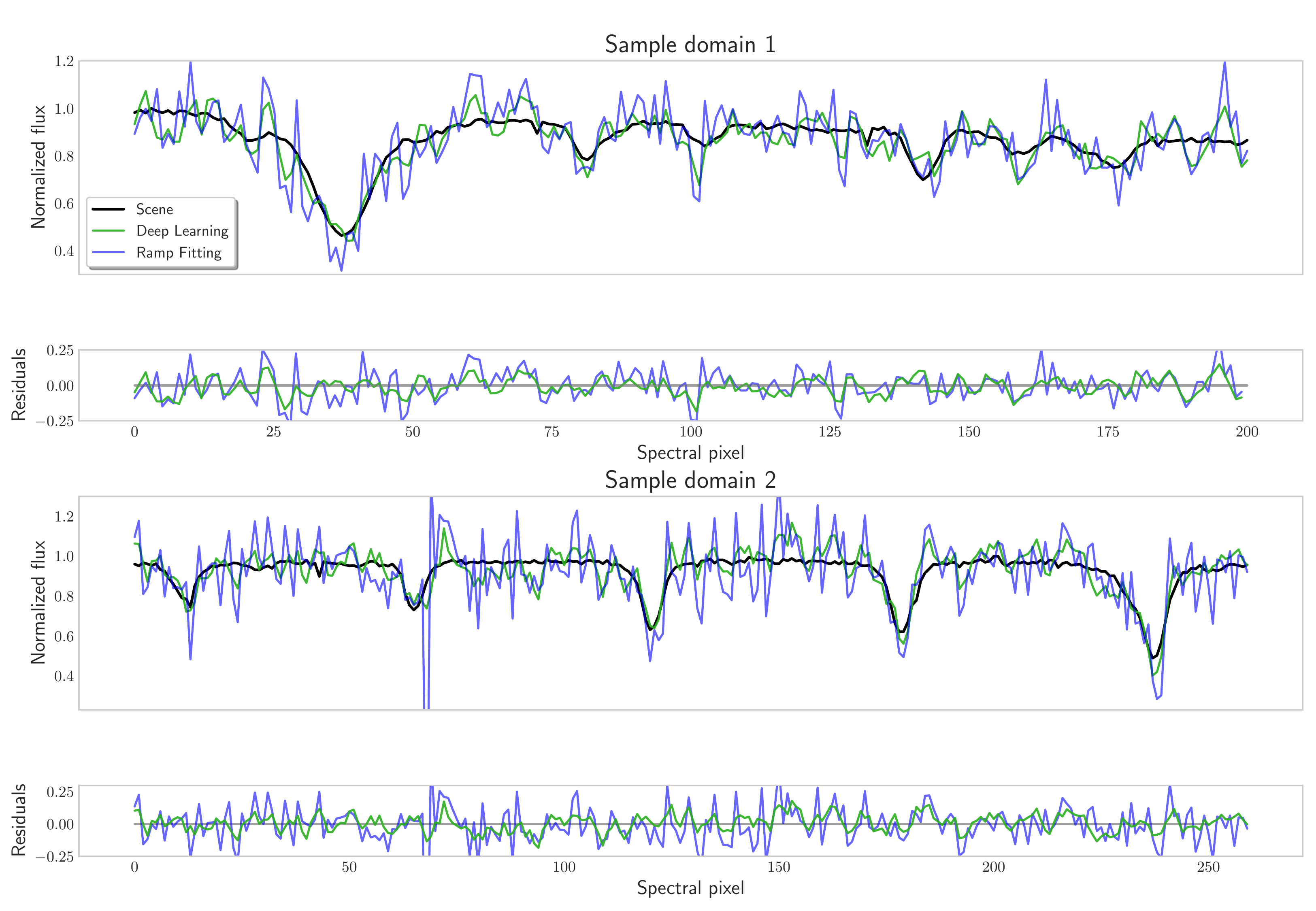}
  \end{center}
  \caption{Extracted spectra on portions of the first test image. Compared are the true spectrum and the extracted spectrum using the neural network and using ramp fitting. The spectra obtained using the neural network are considerably less noisy than the extraction done using ramp fitting, and they do not show any systematic bias. On the top and bottom panels, the standard error to the scene is reduced by a factor of 1.73 and 1.55, respectively , showing that the reduction in per-pixel error has translated into a reduction of error at the spectrum level.}
  \label{fig:Spectrum Extraction good}
\end{figure*}

\begin{figure*}
  \begin{center}
    \includegraphics[width=1\textwidth]{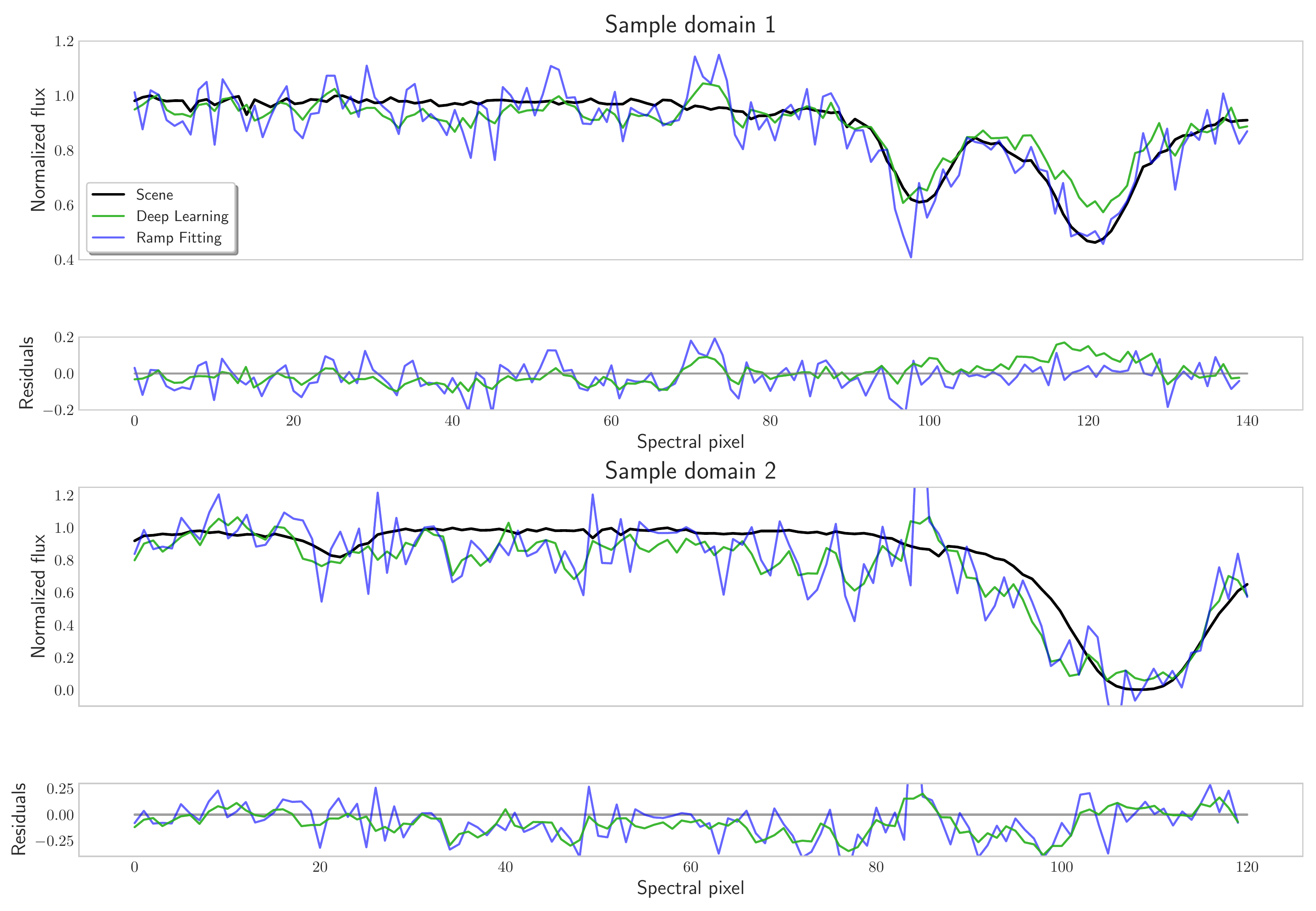}
  \end{center}
  \caption{Presence of biases in spectrum extractions on portions of the first test image. It compares the true spectrum to the spectrum extracted using the neural network and using ramp fitting. The two panels represent the worst case scenario in terms of systematic bias in the spectrum obtained using the neural network.}
  \label{fig:Spectrum Extraction bad}
\end{figure*}

A more telling metric by which to evaluate the performance of the neural network is a spectrum extraction. We can extract the spectrum from science images obtained using the neural network by summing pixels in the spatial direction. Figure \ref{fig:Spectrum Extraction good} presents a small section of the spectrum of the first test image, and compares it to the spectra obtained from the predictions of the neural network, and from ramp fitting. The spectrum obtained from the predictions of the neural network is significantly closer to the truth, and the error reduction over the entirety of the spectra is reported in table \ref{tab: Errors}. Averaged over the three test images, the spectrum-level error is reduced by a factor of 1.85. The neural network performs well on most of the spectrum, but certain sections of the image are problematic, and the spectrum extracted using the neural network contains significant systematic biases (see Figure \ref{fig:Spectrum Extraction bad}). These issues are discussed in more detail in section \ref{sec:Discussion and Future directions}

\subsection{Results on Ten Realisations of a Unique Scene}

\begin{figure*}
  \begin{center}
    \includegraphics[width=1\textwidth]{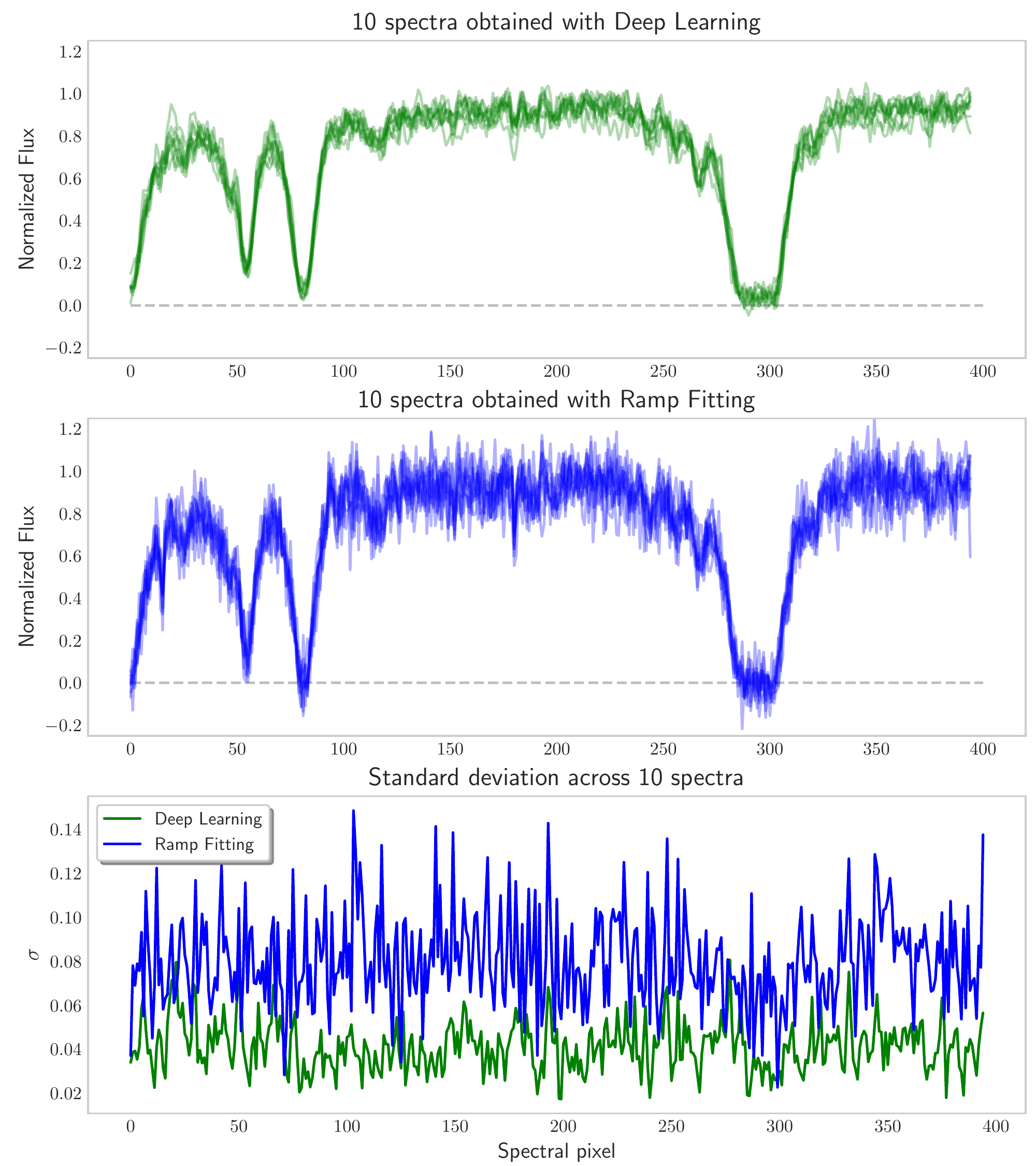}
  \end{center}
  \caption{Small section of spectra extracted for 10 realisations of a unique astrophysical scene, using both the neural network and ramp fitting. We compare the standard deviation $\sigma$ of the 10 spectra. Using the neural network, $\sigma$ is reduced by a factor of 1.93 when averaged over all spectral pixels, signifying that the spectra extracted using the neural network have less variation.}
  \label{fig:10 ramps}
\end{figure*}

As an additional way to quantify the improvement in the error on the spectra obtained using our deep learning method in comparison to the traditional ramp fit method, we create ten realisations of a unique scene. That is, we select a scene and generate ten simulated data ramps using a different laboratory dark ramp each time. For both methods, we extract a small section of the spectrum for each of the ten ramps, and compute the standard deviation of the ten obtained normalized flux values, which we plot as a function of the spectral pixel (see Figure \ref{fig:10 ramps}). For both techniques, we average the standard deviation over all spectral pixels, and obtain that the neural network has lowered the standard deviation by a factor of 1.93.

\subsection{Results on Fabry-Perot Data}

\begin{figure}
  \begin{center}
    \includegraphics[width=0.47\textwidth]{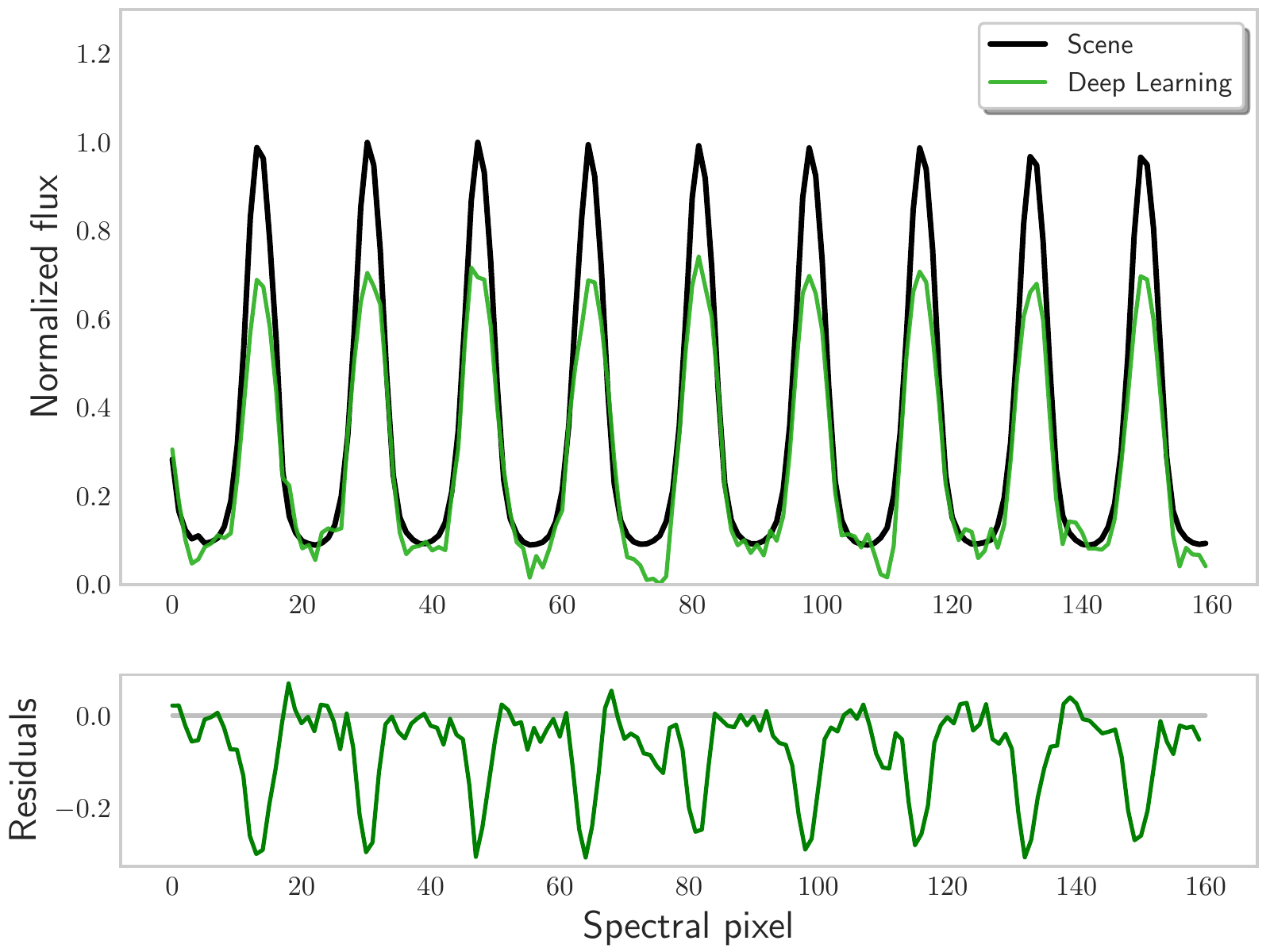}
  \end{center}
  \caption{Portion of spectrum for a Fabry-Perot science image extracted using the neural network. The acceptable accuracy of the extracted spectrum confirms that the neural network is learning statistical properties of the time-correlated noise rather than memorizing and recreating spectra features it has seen during training.}
  \label{fig: freq comb}
\end{figure}

Machine learning models often fail or face difficulties when it comes to generalizations far outside of their training data. In order to test the robustness of the neural network we trained, and to some level its capacity to generalize to scenes that are far from the ones it has seen during training, we use real data collected in the lab using a calibration Fabry-Perot etalon. Figure (\ref{fig: freq comb}) presents a small portion of the spectrum extracted using the neural network. Despite the presence of systematic biases, the fact that the neural network can reconstruct this image with acceptable accuracy despite not having been trained using data resembling this image highlights that it can adapt to types of spectral structures it has never seen. This shows that the neural network is not simply memorising and recreating the absorption lines it sees in training, it is learning the statistical properties of the time-correlated noise present in the detector in a way that allows it so subtract it even when the scene is completely different.

Despite this, in practice, best practices dictate that for science applications, one should use training data as representative as possible of the test data, in order to avoid biases in the predictions due to extrapolation. In fact, it is better to include a wider range of parameters in the training set than what the network is expected to encounter at test time, and then correct for the implicit wide interim prior introduced by this choice of training set through the procedure of \cite{Sebastian2021}, in particular when hierarchical inference of population parameters are of interest.  This section should therefore only be interpreted as a robustness test, which points to the fact that the neural network has indeed learned the statistics of the correlated noise in order to remove it, as we intended. For this reason, we do not provide a comparison between the results obtained using the neural network and ramp fitting on these images.

\section{Discussions and Future directions}
\label{sec:Discussion and Future directions}

Section \ref{sec:Results} demonstrated the neural network presented in this work enables the reduction of readout noise on science images beyond a $1/\sqrt{N}$ decay. 
The fact that our trained neural network returns a more accurate prediction for the flux on a single given pixel than what is theoretically possible using the ramp fit method can be explained by the fact that it uses information that the ramp fit method doesn't: spatial correlations in the scenes. For example, through it's training, the neural network can learn that neighboring pixels on the scenes should have similar fluxes. While this appears to decrease the pixel-by-pixel error on the constructed science images, it also means that the residual error in the neural network pixel flux predictions is correlated across a given forward pass (this therefore represents a correlation of the residual errors of the neural network predictions in the spatial dimension). Because of this correlated error, the lower per-pixel error does not necessarily translate directly into an improvement of the spectrum which can be extracted from the scenes. For this reason, we decide not to use the pixel-level reduction of error on constructed science images to evaluate the performance of the neural network. 

In Section \ref{sec:Results}, we also highlighted the fact that spectra reconstructed from sciences images predicted by our neural network can be more accurate. This represents a better metric by which to quantify the improvement provided by our method in comparison to the traditional ramp fit method. 

However, as shown in Figure \ref{fig:Spectrum Extraction bad}, the spectrum extractions of the neural network sometimes contain systematic errors. Through our tests, we concluded that this occurs when time-correlated noise with frequencies longer than the read time of the three rows given as inputs to the neural network, but shorter than two successive reads on the data ramp. In particular, the neural network tends to produce biased outputs when all rows given to it as inputs are affected similarly by such long-frequency noise, which should be expected as it lacks information on time scales long enough to characterize the properties of such noise. Noise correlated on time scales longer than the time interval input to the network cannot be removed by the neural network. Note that, as shown in Figure \ref{fig:Spectrum Extraction bad}, the ramp-fitting method is similarly affected by such low-frequency noise, and therefore also produces similarly biased output. For this reason, we expect that the neural network could be used in pair with new readout noise removal techniques, especially those that remove correlated noise of lower frequencies. We expect that one could pre-process data through one such algorithm before using the neural network presented here to remove higher-frequency noise. 

To further improve the performance of the neural network, a possibility is the use of truncated back-propagation through time (TBPTT), which would effectively allow inputting a complete data cube to  the neural network, as opposed to a group of three successive rows. In TBPTT, while the updates to the parameters of the GRU cell are still done using a small subsequence of the data due to computational or memory limitations, the memory state of the GRU cell is never reset. Instead, the last hidden state of a subsequence is used as the first hidden state of the next subsequence. Therefore, when a neural network is processing data from a given row of a data cube, it can use, indirectly, information from other rows of the data cube, which may translate into better performance.

Alternatively, we expect that Transformers (see \cite{vaswani_attention_2017}) could act similarly to the GRU-based neural network we developed. Therefore, exploring their performances on this problem could yield interesting results.

In the meantime, we expect that the methods presented in this paper could be used hand in hand with the ramp fit method. In this context, the neural network would act as a way to increase the statistical significance of low certainty detections made using the ramp fit method.

\section{Conclusions}
\label{sec:Conclusions}

In this paper, we presented a new method for the construction of science images from data collected by nIR arrays on HxRG detectors, relying on the use of a neural network. From our testing of the neural network we draw the following conclusions:
\begin{itemize}
    \item The use of time correlations in the data and spatial correlations in the images, enabled by neural networks, allows reaching levels of RON lower than the theoretical $1/\sqrt{N}$ bound for ramp fitting.
    \item In the readout-noise limited regime, the lower level of RON achieved translates into a significant improvement on spectrum extractions done on the constructed images.
    \item We expect that the methods presented here could work hand in hand with either the traditional ramp fit method, or other new readout noise removal methods, and help confirm observations with very little certainty due to a small signal to noise ratio.
    \item This paper represents a proof of concept specific to H4RG detectors and NIRPS, but, we expect that our approach is promising for other applications like SPIRou and JWST.

\end{itemize}

\acknowledgments
We would like to thank Compute Canada (www.computecanada.ca) and Calcul Qu\'{e}bec (www.calculquebec.ca) for providing computational resources and support that contributed to this work. This research was enabled in part by support provided by the Natural Sciences and Engineering Research Council of Canada (NSERC) (grant ID RGPIN-2020-05102). We thank the anonymous referee for their feedback and comments that resulted in a more thorough presentation  of our methods.

\appendix

\section{Theoretical effective RON decrease from multiple readouts in the absence of time-correlated noise}
\label{appendix}

We determine the theoretical rate at which readout noise falls with the number of readouts in a ramp, N, in the event that no time-correlated noise is present in the data, and ramp fitting is used to generate the predictions. 

So that an analytic equation may be obtained, we first consider the limit case where there is no flux anywhere on the array. Then, the only source of noise is detector noise, which we assume is Gaussian with mean $\mu=0$ and standard deviation $\sigma$. The data collected for any pixel on the array is a collection of $N$ samples from the instrument noise. ramp fitting consists of placing these samples at the $x $ coordinates $(1,\dots,N)$ and finding the least-squares line going through the data. 

Let $F$ be the prediction for the flux on a pixel. Since the true flux is 0, the expected error on $F$ is given by Std$(F)$. This is the quantity to be computed as a function of $N$. 

Define the matrices
\begin{align*}
    X=\begin{bmatrix}
    1 & 1\\
    1 & 2\\
    \vdots & \vdots\\
    1 & N\\
    \end{bmatrix}, Y=\begin{bmatrix}
    y_1\\
    y_2\\
    \vdots\\
    y_N
    \end{bmatrix}, B=\begin{bmatrix}
    b\\
    m\\
    \end{bmatrix},
\end{align*}
where $y_i$, $i=(1,\dots,N)$ are sampled from a Gaussian distribution with mean $\mu =0$, standard deviation $\sigma$, and $B$ is the matrix of best-fit parameters. We first derive an equation for $B$. Using the prime symbol as notation for the transpose of a matrix, we have a chi-squared given by
\begin{align}
    \chi^2 = \sum_{i=1}^{N}\frac{(y_i-(mx_i+b))^2}{\sigma^2} = \frac{1}{\sigma^2}(Y-XB)'(Y-XB).
\end{align}
The best-fit parameters are those that minimise the chi-squared. So we take it's derivative with respect to $B$, set it to 0, and solve for $B$. That gives
\begin{align}
    \pdv{\chi^2}{B} &= \pdv{B}\Big(\frac{1}{\sigma^2}(Y-XB)'(Y-XB)\Big) \eqset 0\\
    &\implies \frac{1}{\sigma^2}\Big(\pdv{(Y-XB)'}{B} \ (Y-XB) + (Y-XB)'\pdv{(Y-XB)}{B}\Big) = 0\\
    &\implies \Big(-X'(Y-XB) + (Y-XB)'(-X)\Big) = 0\\
    &\implies X'(Y-XB) = 0\\
    &\implies B = (X'X)^{-1}X'Y \label{ref1}.
\end{align}
We now turn to finding the covariance matrix $\Cov(B)$ of $B$. It is given by
\begin{align}
    \Cov(B) &= \E[(B-\E[B])^2]\\
    &= \E[BB'-2B\E(B)+E[B]^2]\\
    &= \E[BB']-\E[B]^2\\
    &= \E[BB'].
\end{align}
where the last equality follows from the fact that $\E(B)$ is the matrix of true parameters, which are all 0 since we have assumed 0 flux. Using (\ref{ref1}), we get
\begin{align}
    \Cov(B) &= \E[((X'X)^{-1}X'Y)((X'X)^{-1}X'Y)']\\
    &= \E[(X'X)^{-1} YY'X(X'X)^{-1}]\\
    &= (X'X)^{-1}X'\E(YY')X(X'X)^{-1}\\
    &= \sigma^2(X'X)^{-1}.
\end{align}
Substituting in the definition of $X$ and carrying out the matrix multiplication gives
\begin{align}
    \Cov(B) &= \sigma^2\begin{bmatrix}
    N & \frac{1}{2}(N)(N+1)\\
    \frac{1}{2}(N)(N+1) & \frac{1}{6}(N)(N+1)(2N+1)
    \end{bmatrix}^{-1}\\
    &= \begin{bmatrix}
    \sigma^2\mfrac{4N+2}{N^2-N} & \sigma^2\mfrac{-6}{N^2-N}\\[1ex]
    \sigma^2\mfrac{-6}{N^2-N} & \sigma^2\mfrac{12}{N^3-N}
    \end{bmatrix}.
\end{align}
The bottom right entry is the variance on $m$. Taking its square root gives
\begin{align}
    \text{Std}(m) = \frac{\sqrt{12}\sigma}{\sqrt{N^3-N}} \label{ref2}.
\end{align}
To find the standard error on the predicted flux $F$, we multiply what we got in (\ref{ref2}) by $(N-1)$. That gives
\begin{align}
    \text{Std}(F) &= \sqrt{12}\sigma\frac{N-1}{\sqrt{N^3-N}}\\
    &= \sqrt{12}\sigma\sqrt{\frac{N-1}{(N)(N+1)}}, \label{ref3}
\end{align}
which for sufficiently large $N$ is well approximated by
\begin{align}
    \text{Std}(F) = \sqrt{12}\sigma\frac{1}{\sqrt{N}}.
\end{align}
The consequence of this is that readout noise falls with $1/\sqrt{N}$ as expected.

\begin{figure}[H]
  \begin{center}
    \includegraphics[width=0.7\textwidth]{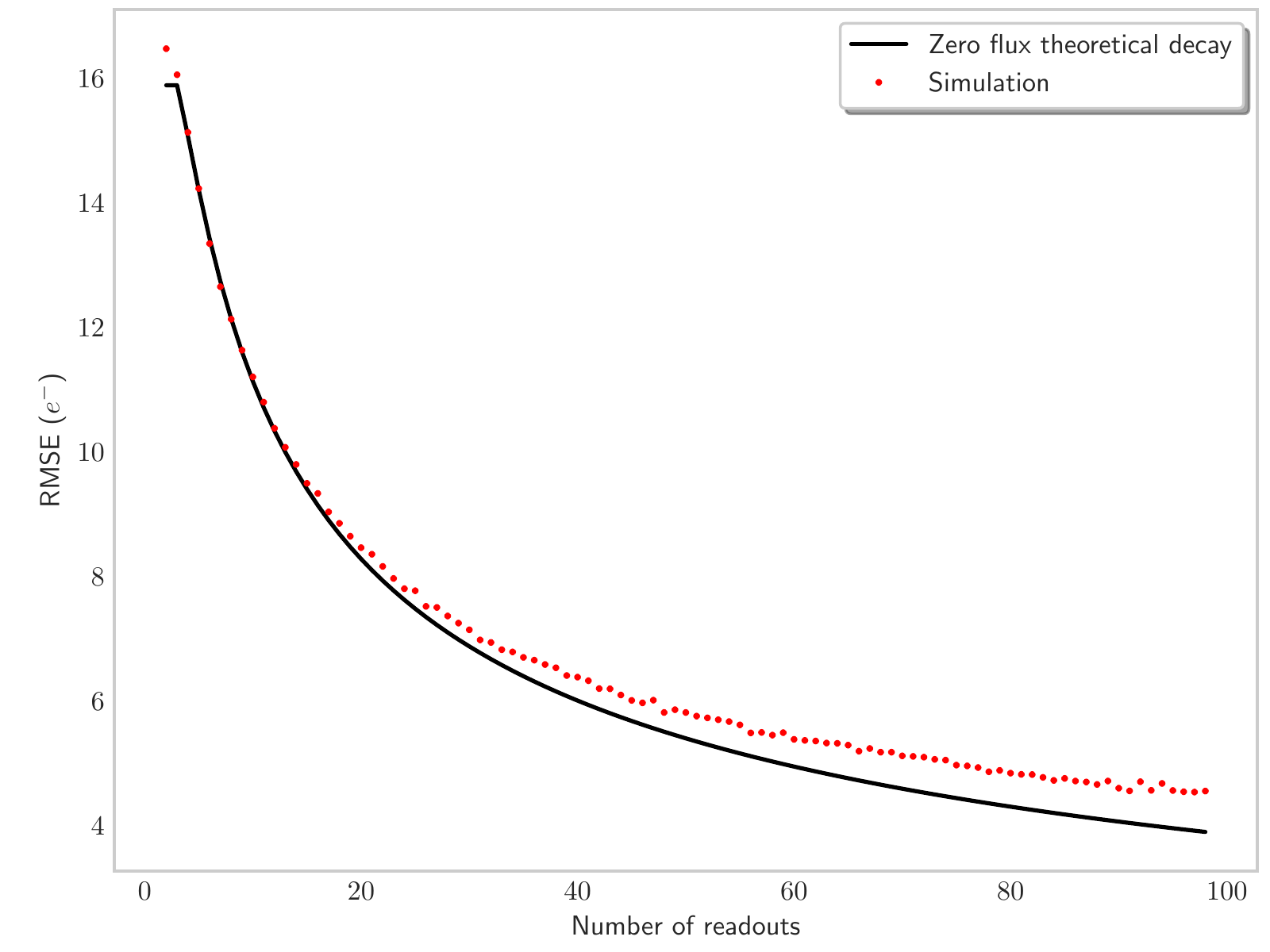}
  \end{center}
  \caption{Example of a simulation of the RON decay in the case of no correlated noise vs. the theoretical decay for an array receiving no flux. The difference between the two decays is small.}
  \label{fig:RON simulation}
\end{figure}

When flux is added to the image, analytic calculations are harder and impractical, because photon noise is not noise around a slope, it is the result of a Poisson accumulation. We opt for a simulation of the RON decay instead. We use a scene used during training of the neural network to get a realistic distribution of flux values on the array. We generate a Poisson accumulation for each of these pixels. We add instrument noise to each readout, which we again assume is Gaussian with mean $\mu = 0$ and standard deviation $\sigma$. We fit a least square line through the generated data and compare the predictions for the flux to the true values. We repeat this process for multiple values of $N$ in order to get a portrait of the RON fallout. The observation is that the RON decay is only slightly slower than the theoretical decay for an array receiving no flux derived in equation (\ref{ref3}) (see Figure \ref{fig:RON simulation}). Over 100 readouts, the difference is small. This shows that the observed RON decay on data cubes collected by HxRG detectors in low-flux regimes can be improved significantly via the removal of correlated readout noise.

\newpage
\bibliography{references.bib}

\begin{thebibliography}{}
\expandafter\ifx\csname natexlab\endcsname\relax\def\natexlab#1{#1}\fi
\providecommand{\url}[1]{\href{#1}{#1}}
\providecommand{\dodoi}[1]{doi:~\href{http://doi.org/#1}{\nolinkurl{#1}}}
\providecommand{\doeprint}[1]{\href{http://ascl.net/#1}{\nolinkurl{http://ascl.net/#1}}}
\providecommand{\doarXiv}[1]{\href{https://arxiv.org/abs/#1}{\nolinkurl{https://arxiv.org/abs/#1}}}

\bibitem[{Abadi {et~al.}(2015)Abadi, Agarwal, Barham, Brevdo, Chen, Citro,
  Corrado, Davis, Dean, Devin, Ghemawat, Goodfellow, Harp, Irving, Isard, Jia,
  Jozefowicz, Kaiser, Kudlur, Levenberg, Man\'{e}, Monga, Moore, Murray, Olah,
  Schuster, Shlens, Steiner, Sutskever, Talwar, Tucker, Vanhoucke, Vasudevan,
  Vi\'{e}gas, Vinyals, Warden, Wattenberg, Wicke, Yu, \&
  Zheng}]{tensorflow2015-whitepaper}
Abadi, M., Agarwal, A., Barham, P., {et~al.} 2015, {TensorFlow}: Large-Scale
  Machine Learning on Heterogeneous Systems.
\newblock \url{https://www.tensorflow.org/}

\bibitem[{Baso {et~al.}(2019)Baso, Rodríguez, \& Danilovic}]{baso_solar_2019}
Baso, C. J.~D., Rodríguez, J. d. l.~C., \& Danilovic, S. 2019, Astronomy \&
  Astrophysics, 629, A99, \dodoi{10.1051/0004-6361/201936069}

\bibitem[{{Birkmann} {et~al.}(2018){Birkmann}, {Sirianni}, {Ferruit},
  {Willott}, {Maiolino}, {Rauscher}, {Alves de Oliveira}, {B{\"o}ker},
  {Giardino}, {L{\"u}tzgendorf}, {Marston}, {Puga}, {Rawle}, {te Plate},
  {Jensen}, \& {Rumler}}]{Birkmann2018}
{Birkmann}, S.~M., {Sirianni}, M., {Ferruit}, P., {et~al.} 2018, in Society of
  Photo-Optical Instrumentation Engineers (SPIE) Conference Series, Vol. 10709,
  High Energy, Optical, and Infrared Detectors for Astronomy VIII, ed. A.~D.
  {Holland} \& J.~{Beletic}, 1070930, \dodoi{10.1117/12.2313524}

\bibitem[{Bouchy {et~al.}(2017)Bouchy, Doyon, Artigau, Melo, Hernandez, Wildi,
  Delfosse, Lovis, Figueira, Canto~Martins, González~Hernández, Thibault,
  Reshetov, Pepe, Santos, De~Medeiros, Rebolo, Abreu, Adibekyan, Bandy, Benz,
  Blind, Bohlender, Boisse, Bovay, Broeg, Brousseau, Cabral, Chazelas,
  Cloutier, Coelho, Conod, Cumming, Delabre, Genolet, Hagelberg, Jayawardhana,
  Käufl, Lafrenière, De~Castro~Leão, Malo, De~Medeiros~Martins, Matthews,
  Metchev, Oshagh, Ouellet, Parro, Rasilla~Piñeiro, Santos, Sarajlic, Segovia,
  Sordet, Udry, Valencia, Vallée, Venn, Wade, \&
  Saddlemyer}]{bouchy_near-infrared_2017}
Bouchy, F., Doyon, R., Artigau, E., {et~al.} 2017, Published in The Messenger
  vol. 169, pp. 21-27, 7 pages, \dodoi{10.18727/0722-6691/5034}

\bibitem[{Chollet {et~al.}(2015)}]{chollet2015keras}
Chollet, F., {et~al.} 2015, Keras,  GitHub.
\newblock \url{https://github.com/fchollet/keras}

\bibitem[{Esteva {et~al.}(2017)Esteva, Kuprel, Novoa, Ko, Swetter, Blau, \&
  Thrun}]{esteva_dermatologist-level_2017}
Esteva, A., Kuprel, B., Novoa, R.~A., {et~al.} 2017, Nature, 542, 115,
  \dodoi{10.1038/nature21056}

\bibitem[{Fowler \& Gatley(1990)}]{fowler_demonstration_1990}
Fowler, A.~M., \& Gatley, I. 1990, The Astrophysical Journal, 353, L33,
  \dodoi{10.1086/185701}

\bibitem[{{Gardner} {et~al.}(2006){Gardner}, {Mather}, {Clampin}, {Doyon},
  {Greenhouse}, {Hammel}, {Hutchings}, {Jakobsen}, {Lilly}, {Long}, {Lunine},
  {McCaughrean}, {Mountain}, {Nella}, {Rieke}, {Rieke}, {Rix}, {Smith},
  {Sonneborn}, {Stiavelli}, {Stockman}, {Windhorst}, \& {Wright}}]{Gardner2006}
{Gardner}, J.~P., {Mather}, J.~C., {Clampin}, M., {et~al.} 2006, \ssr, 123,
  485, \dodoi{10.1007/s11214-006-8315-7}

\bibitem[{{Hall} {et~al.}(2012){Hall}, {Atkinson}, {Beletic}, {Blank},
  {Farris}, {Hodapp}, {Jacobson}, {Loose}, \& {Luppino}}]{Hall2012}
{Hall}, D. N.~B., {Atkinson}, D., {Beletic}, J.~W., {et~al.} 2012, in Society
  of Photo-Optical Instrumentation Engineers (SPIE) Conference Series, Vol.
  8453, High Energy, Optical, and Infrared Detectors for Astronomy V, ed. A.~D.
  {Holland} \& J.~W. {Beletic}, 84530W, \dodoi{10.1117/12.927226}

\bibitem[{{Hezaveh} {et~al.}(2017){Hezaveh}, {Perreault Levasseur}, \&
  {Marshall}}]{Hezaveh2017}
{Hezaveh}, Y.~D., {Perreault Levasseur}, L., \& {Marshall}, P.~J. 2017, \nat,
  548, 555, \dodoi{10.1038/nature23463}

\bibitem[{Hložek(2019)}]{hlozek_data_2019}
Hložek, R. 2019, Publications of the Astronomical Society of the Pacific, 131,
  118001, \dodoi{10.1088/1538-3873/ab311d}

\bibitem[{Husser {et~al.}(2013)Husser, Wende-von Berg, Dreizler, Homeier,
  Reiners, Barman, \& Hauschildt}]{husser_new_2013}
Husser, T.-O., Wende-von Berg, S., Dreizler, S., {et~al.} 2013, Astronomy \&
  Astrophysics, 553, A6, \dodoi{10.1051/0004-6361/201219058}

\bibitem[{Kingma \& Ba(2017)}]{kingma_adam_2017}
Kingma, D.~P., \& Ba, J. 2017, arXiv:1412.6980 [cs].
\newblock \url{http://arxiv.org/abs/1412.6980}

\bibitem[{{Li} {et~al.}(2021){Li}, {Ni}, {Croft}, {Di Matteo}, {Bird}, \&
  {Feng}}]{YinLiPNAS2021}
{Li}, Y., {Ni}, Y., {Croft}, R. A.~C., {et~al.} 2021, Proceedings of the
  National Academy of Science, 118, 2022038118, \dodoi{10.1073/pnas.2022038118}

\bibitem[{{Mosby} {et~al.}(2020){Mosby}, {Rauscher}, {Bennett}, {Cheng},
  {Cheung}, {Cillis}, {Content}, {Cottingham}, {Foltz}, {Gygax}, {Hill},
  {Kruk}, {Mah}, {Meier}, {Merchant}, {Miko}, {Piquette}, {Waczynski}, \&
  {Wen}}]{Mosby2020}
{Mosby}, G., {Rauscher}, B.~J., {Bennett}, C., {et~al.} 2020, Journal of
  Astronomical Telescopes, Instruments, and Systems, 6, 046001,
  \dodoi{10.1117/1.JATIS.6.4.046001}

\bibitem[{{Ni} {et~al.}(2021){Ni}, {Li}, {Lachance}, {Croft}, {Di Matteo},
  {Bird}, \& {Feng}}]{Ni2021}
{Ni}, Y., {Li}, Y., {Lachance}, P., {et~al.} 2021, \mnras, 507, 1021,
  \dodoi{10.1093/mnras/stab2113}

\bibitem[{Rauscher {et~al.}(2017)Rauscher, Arendt, Fixsen, Greenhouse, Lander,
  Lindler, Loose, Moseley, Mott, Wen, Wilson, \&
  Xenophontos}]{rauscher_improved_2017}
Rauscher, B.~J., Arendt, R.~G., Fixsen, D.~J., {et~al.} 2017, Publications of
  the Astronomical Society of the Pacific, 129, 105003,
  \dodoi{10.1088/1538-3873/aa83fd}

\bibitem[{Siemiginowska {et~al.}(2019)Siemiginowska, Eadie, Czekala, Feigelson,
  Ford, Kashyap, Kuhn, Loredo, Ntampaka, Stevens, Avelino, Borne, Budavari,
  Burkhart, Cisewski-Kehe, Civano, Chilingarian, van Dyk, Fabbiano, Finkbeiner,
  Foreman-Mackey, Freeman, Fruscione, Goodman, Graham, Guenther, Hakkila,
  Hernquist, Huppenkothen, James, Law, Lazio, Lee, López-Morales, Mahabal,
  Mandel, Meng, Moustakas, Muna, Peek, Richards, Portillo, Scargle, de~Souza,
  Speagle, Stassun, Stenning, Taylor, Tremblay, Trimble, Yanamandra-Fisher, \&
  Young}]{siemiginowska_astro2020_2019}
Siemiginowska, A., Eadie, G., Czekala, I., {et~al.} 2019, arXiv:1903.06796
  [astro-ph].
\newblock \url{http://arxiv.org/abs/1903.06796}

\bibitem[{Silver {et~al.}(2017)Silver, Schrittwieser, Simonyan, Antonoglou,
  Huang, Guez, Hubert, Baker, Lai, Bolton, Chen, Lillicrap, Hui, Sifre, van~den
  Driessche, Graepel, \& Hassabis}]{silver_mastering_2017}
Silver, D., Schrittwieser, J., Simonyan, K., {et~al.} 2017, Nature, 550, 354,
  \dodoi{10.1038/nature24270}

\bibitem[{Ulyanov {et~al.}(2020)Ulyanov, Vedaldi, \&
  Lempitsky}]{ulyanov_deep_2020}
Ulyanov, D., Vedaldi, A., \& Lempitsky, V. 2020, International Journal of
  Computer Vision, 128, 1867, \dodoi{10.1007/s11263-020-01303-4}

\bibitem[{Vaswani {et~al.}(2017)Vaswani, Shazeer, Parmar, Uszkoreit, Jones,
  Gomez, Kaiser, \& Polosukhin}]{vaswani_attention_2017}
Vaswani, A., Shazeer, N., Parmar, N., {et~al.} 2017, arXiv:1706.03762 [cs].
\newblock \url{http://arxiv.org/abs/1706.03762}

\bibitem[{Venn {et~al.}(2019)Venn, Fabbro, Liu, Hezaveh, Perreault-Levasseur,
  Eadie, Ellison, Woo, Kavelaars, Yi, Hlozek, Bovy, Teimoorinia, Ravanbakhsh,
  \& Spencer}]{venn_lrp2020_2019}
Venn, K.~A., Fabbro, S., Liu, A., {et~al.} 2019, arXiv:1910.00774 [astro-ph],
  \dodoi{10.5281/zenodo.3755910}

\bibitem[{{Wagner-Carena} {et~al.}(2021){Wagner-Carena}, {Park}, {Birrer},
  {Marshall}, {Roodman}, {Wechsler}, \& {LSST Dark Energy Science
  Collaboration}}]{Sebastian2021}
{Wagner-Carena}, S., {Park}, J.~W., {Birrer}, S., {et~al.} 2021, \apj, 909,
  187, \dodoi{10.3847/1538-4357/abdf59}

\bibitem[{Wei \& Huerta(2020)}]{wei_gravitational_2020}
Wei, W., \& Huerta, E. 2020, Physics Letters B, 800, 135081,
  \dodoi{10.1016/j.physletb.2019.135081}

\bibitem[{Werbos(1990)}]{werbos_backpropagation_1990}
Werbos, P. 1990, Proceedings of the IEEE, 78, 1550, \dodoi{10.1109/5.58337}

\end{thebibliography}
\bibliographystyle{aasjournal}

\end{document}